\documentclass{jfm_old}

\usepackage{graphicx}
\usepackage{newtxtext}
\usepackage{newtxmath}
\usepackage{natbib}
\usepackage{hyperref}
\hypersetup{
    colorlinks = true,
    urlcolor   = blue,
    citecolor  = black,
}

\newcommand{\RomanNumeralCaps}[1]
\linenumbers

\title{A method for evaluating relations of turbulent normal-stresses by experimental data over a wide range of Reynolds numbers}

\author{Hassan Nagib\aff{1}
  \corresp{\email{nagib@iit.edu}},
 \and Ivan Marusic\aff{2}}

\affiliation{\aff{1} ILLINOIS TECH. Chicago, IL 60616, USA
\aff{2}University of Melbourne, Parkville, VIC 3010, Australia}

\begin{document}
\maketitle

\begin{abstract}
Recently, \cite{nag24} utilized indicator functions of profiles of the streamwise normal stress to reveal the ranges of validity, in wall distance and Reynolds number, for each of two proposed models in Direct Numerical Simulations (DNS) of channel and pipe flows. A method more suited to experimental data is proposed here, as establishing accurate indicator functions is a challenge. The new method is outlined and used with the two leading models that propose either a logarithmic or power trend, for the normal stresses of turbulence in a ``fitting region" of wall-bounded flows. The method, which is simple and robust, is used to evaluate each model over a wide range of Reynolds number ($Re_\tau$) by applying it to two of the prominent experimental data sets in zero-pressure-gradient boundary layers (ZPG) and pipe flows. Significant regions of validity by either relation are only found when $Re_\tau \gtrapprox 10,000$, with the lower limit $y^+_{in} \sim (Re_\tau)^{0.5}$ and the provision that $y^+_{in} \gtrapprox 400$ for the ZPG and pipe flow data. The upper limit is found to be a fixed fraction of the boundary layer thickness or the pipe radius, and is independent of $Re_\tau$. However, the outer limit of validity for the power trend is about twice that of the logarithmic trend, and is identified at $Y \approx 0.4$ for ZPG and $Y \approx 0.5$ for pipe flow data. A somewhat larger exponent for the power trend equal to $0.28$ is found to slightly extend the range of validity of the power model, compared to the $0.25$ power of \cite{che22,che23}. Correcting for outer intermittency in ZPG data also extends validity of the power trend to around half the boundary layer thickness. Projecting near-wall peak normal-stress values using both models based on data from locations in the fitting region, yields nearly identical results up to the highest $Re_\tau$ available from these two experiments. Neither model directly represents the inner peak values and differ by as much as $30 \%$. However, when the trends of the models are anchored by experimental data, the two relations provide projected values at $Re_\tau = 10^6$ with a relative difference of approximately $6.2\%$; distinguishing between these values would require measurement accuracy well beyond our capabilities. Current results and those of \cite{mon23} and \cite{bax24} on the mean flow, suggest that considerations of nonlinear growth of eddies from the wall and accounting for some viscous effects, are important for modeling wall-bounded flows. Such effects also apply to potential refinements of the logarithmic trend along ideas advanced by \cite{Desh21}.
\end{abstract}


\newpage

\section{Introduction and data sets used}
Interest in characterizing the normal stress of turbulent fluctuations and its relationship to large-scale eddies has significantly increased  over the last five to ten years. Two models have been proposed to represent the trends for normal stresses of turbulence in wall-bounded flows, and they are the focus of recent research and publications. \cite{nag24} utilized indicator functions with carefully selected Direct Numerical Simulations (DNS) data for channel and pipe flows to examine the validity of both models and their potential applicable ranges. The indicator function approach was also tried for experimental data by \cite{M23}. The current work is also motivated by recent findings regarding an extended overlap region of the mean velocity profile for wall-bounded turbulent flows by \cite{mon23}, revealing that not all such flows exhibit a pure logarithmic profile, and a linear term of the same order should be considered. One potential implication from the additional linear term in the mean flow is that it would require eddies that do not strictly scale with their distance from the wall in this overlap region. 

One model is based on the attached-eddy concept described by \cite{mar19}. Here, we will refer to this model as the ``wall-scaled eddies model" or simply the ``logarithmic trend" model.
For the streamwise normal stress, the trend based on this model is given by:

\begin{equation}\label{eq:001}
     \left\langle u^+u^+\right\rangle (Y) = B_1 - A_1\ln(Y).
\end{equation}

The outer-scaled distance from the wall $Y$ is defined as $y/\delta$ in the ZPG data and as $y/R$ for the pipe flow, where $\delta$ is the boundary layer thickness and $R$ the pipe radius. The $\delta$ used here is the same as that used by \cite{sam18} and is equivalent to the one defined by \cite{bax24}.

Recent publications by \cite{che22,che23} on bounded dissipation, introduce an alternative model, which we will refer to as the ``power trend" model. This model is represented by the following relation for the streamwise normal stress:

\begin{equation}\label{eq:002}
    \left\langle u^+u^+\right\rangle (Y) = \alpha_1 - \beta_1~(Y)^{0.25}.
\end{equation}

Both models contain two parameters, and in addition the power trend exponent may also be considered as a potential third parameter. The subscript ``$1$" represents the streamwise component of the normal stresses of turbulence.  Agreement with the logarithmic trend is often evaluated in the literature, particularly with comparison to experimental data, by fitting a straight line over a segment of the normal-stress data on a semi-log plot; examples are found in \cite{mar13}, and more recently in \cite{hwa22} and \cite{Diw21}.  This approach makes it challenging to assess the accuracy and ranges of validity of the models, particularly due to the often sparse logarithmic spacing of results in the fitting and outer regions, as well as the limited accuracy of experimental data. For the $0.25$-power trend, agreement with data is typically tested by iteratively adjusting the two parameters $\alpha_1$ and $\beta_1$, seeking minimum deviations over widest possible range of distances from the wall in a fitting region between wall- and outer-flow parts.

In the case of DNS data, indicator functions analogous to those utilized with  mean velocity profiles in \cite{mon23} were identified as having the best potential for normal stress analysis, and were favored by \cite{nag24}. However, just like with mean velocity, these indicator functions require taking derivatives of the discrete normal-stress profiles.  This approach is often unsuitable for most experimental data in the literature due to limited spatial resolution and experimental accuracy, which hamper the ability to obtain accurate derivatives of the profiles.  The current work aims to develop a method for assessing the proposed relations using two of the most well-documented and reliable experimental results in zero-pressure-gradient boundary layers and pipe flows, as provided by \cite{sam18} and \cite{hul12}, respectively.

A new approach is introduced here and successfully implemented with these two experimental data sets over a ``fitting region" defined for this work by the following. It is recommended for use only with accurate measurements with established uncertainty limits. The method is based on a normalization scheme that utilizes the respective trend relations with adjustable parameters within defined bounds.  The bounds are selected to establish a fitting region for each model between inner and outer flow parts.  To meet this criterion, the current assessment of the streamwise normal stress primarily focuses on regions of the flow beyond the very near wall,  specifically targeting ranges of $y^+$ larger than 400, which correspond to a nominal outer variable range in case of ZPG of $0.004 \lessapprox Y \lessapprox 0.2$ for the logarithmic trend and $0.004 \lessapprox Y \lessapprox 0.4$ for the power trend in the $Re_\tau$ range examined. For the logarithmic trend of normal stresses, the overlap region is not expected to extend beyond $Y$ of $0.1$ or $0.15$. The inner-scaled wall distance is defined as $y^+ = y u_\tau/\nu$, where $Re_\tau$ is equal to $u_\tau \delta/\nu$ for ZPG flows and $u_\tau R/\nu$ for pipes. The fluid kinematic viscosity $\nu$ is determined from the fluid temperature and $u_\tau$ is the wall friction velocity $\sqrt{\tau_w/\rho}$, with $\tau_w$ measured directly using oil-film interferometry in the ZPG experiment and by pressure drop in the fully developed pipe flow.

It is important to highlight that both models are intended to represent regions between the near-wall flow and the ``outer" part of the flow. For the different wall-bounded turbulent flows, the outer part of the flow can differ a great deal from boundary layers to duct flows such as in channels and pipes. We hope that the approach described and tested in this work will reveal some differences among these fitting regions, although the near-wall flow may be quite similar. For these high $Re_\tau$ experiments, the fitting region is selected to incorporate the majority of the inner part of the overlap region between inner and outer flow parts as defined by each model, while extending to wall distances in outer variable $Y$ to include the range of agreement between the data and each model.

Initial evaluations for the power trend model were made with the exponent of $0.25$ based on the bounded dissipation results.  We also tested other exponents in the range from $0.2$ to $0.32$. For ZPG boundary layers, data from the Melbourne large wind tunnel by \cite{sam18} and \cite{mar15} were used, with emphasis on the recent results utilizing more advanced hot wire anemometers that provided better spatial resolution. Similarly for the normal-stress data from the Princeton Superpipe facility, the more recent data by NSTAP hot-wire sensors with a smaller sensor length from \cite{hul12} were utilized.

For the DNS results, the various channel and pipe flows data recently examined with the indicator functions approach by \cite{nag24} are used to re-evaluate both trend models by the new method. These results were also compared to the results for higher $Re_\tau$ from the two experiments examined in the next section.

\section{Evaluating logarithmic and quarter-power relations}
\begin{figure}
      \centering
      \centering
      \includegraphics[width=0.95\textwidth]{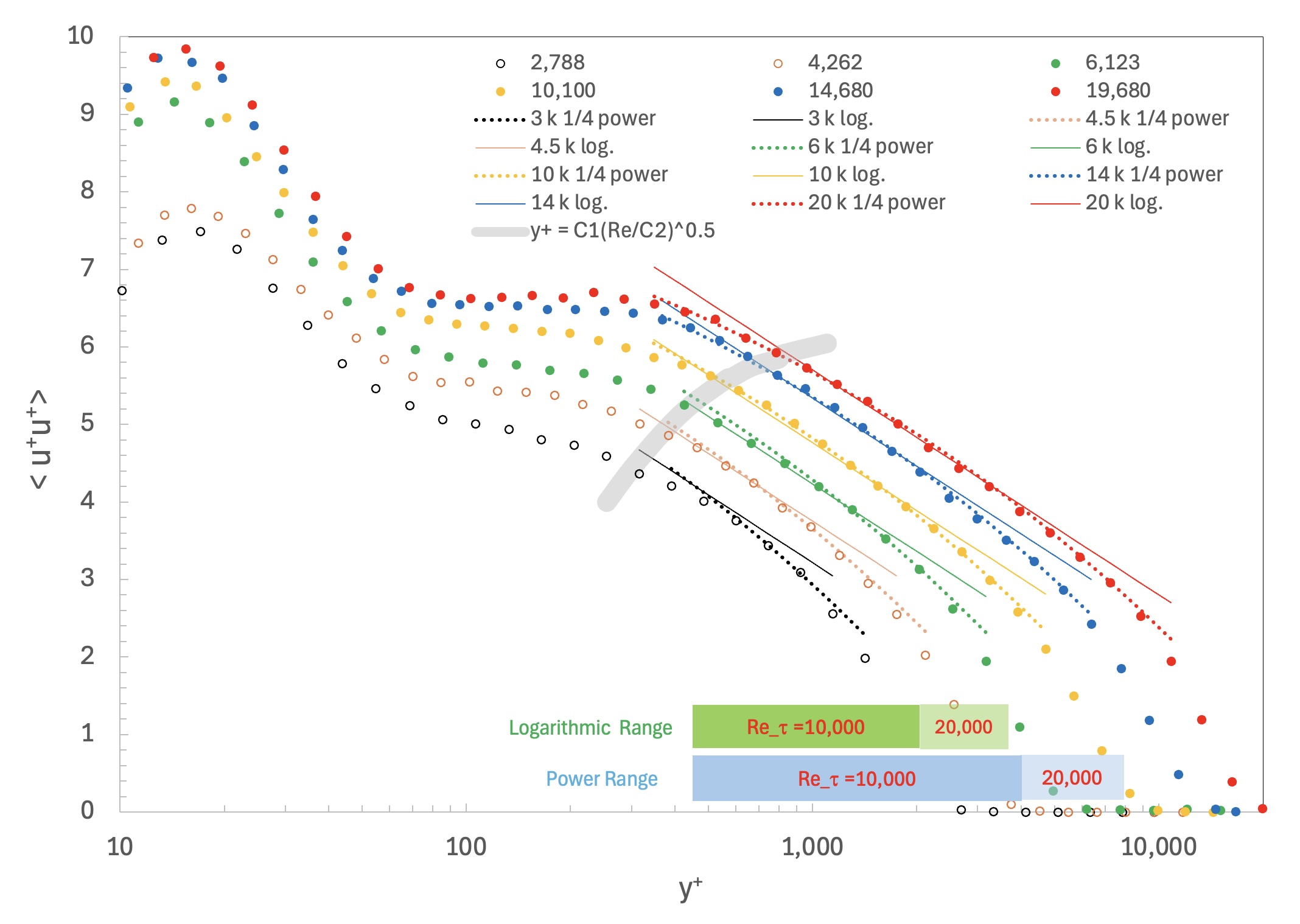}
    \caption{\label{fig:fig1} Streamwise turbulence stress versus inner-scaled wall distance for different $Re_\tau$'s from \cite{sam18} and \cite{mar15} ZPG (open symbols) data with 0.25-power parameters (-9.2 $\pm$ 0.22, 10.1$\pm$ 0.13) and logarithmic parameters (-1.26, 1.93$\pm$ 0.05); C1 = 400 and C2 = 5,000.}
     
\end{figure}

\begin{figure}
       \centering
      \includegraphics[width=0.95\textwidth]{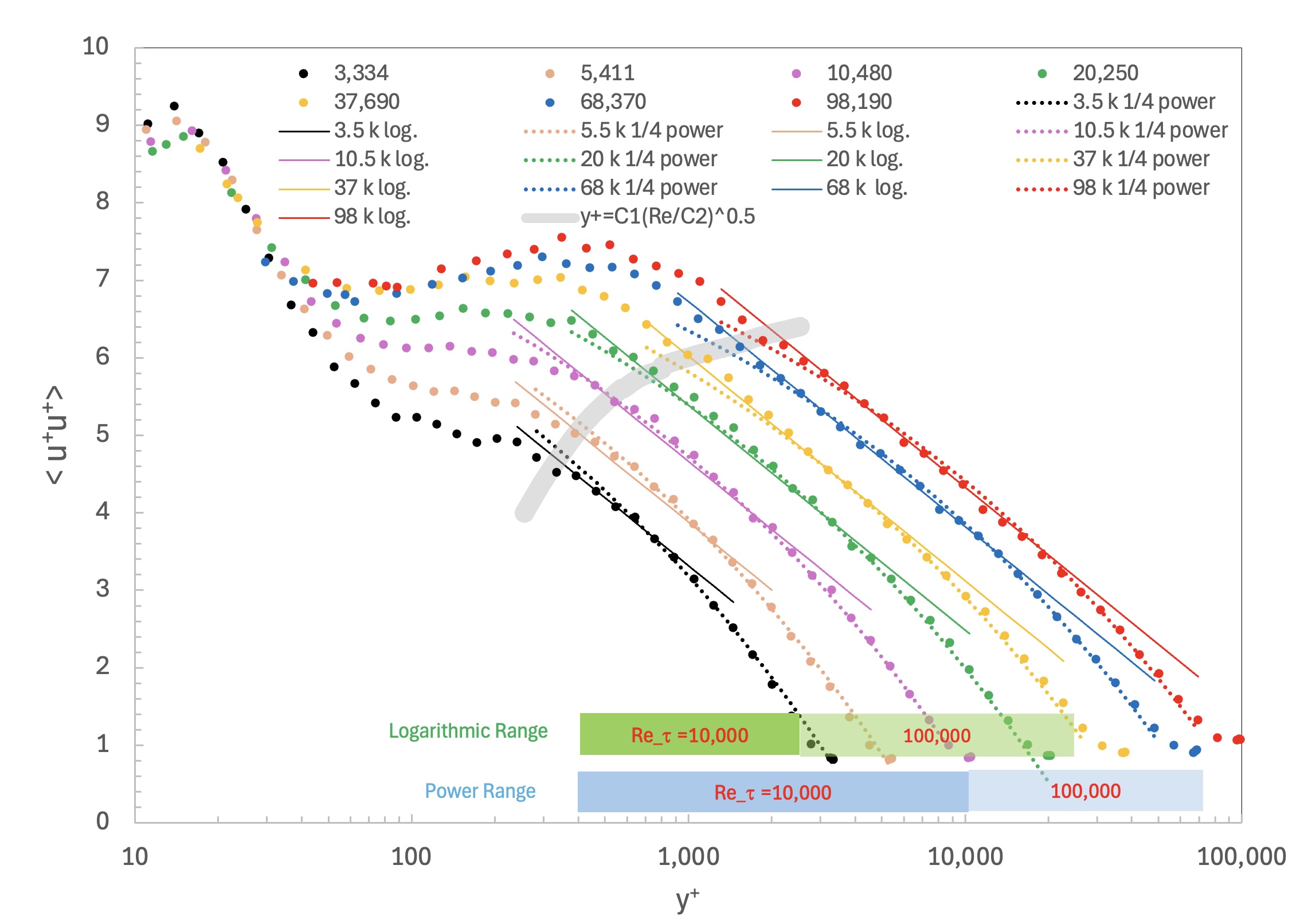}
    \caption{\label{fig:fig2} Streamwise turbulence stress versus inner-scaled wall distance for different $Re_\tau$'s from \cite{hul12} Superpipe data with 0.25-power parameters (-9.3 +/- 0.13, 10.0 $\pm$ 0.0.06), and logarithmic parameters (-1.26, 1.6 $\pm$ 0.17); C1 = 400 and C2 = 5,000.}
     
\end{figure}

The collection of the streamwise normal-stress profiles from the Melbourne ZPG experiments and the Princeton Superpipe are plotted against the logarithm of the inner-scale wall distance in figures~\ref{fig:fig1} and \ref{fig:fig2}, respectively. The logarithmic trend is represented by solid straight lines using equation \ref{eq:001} with typical $A_1$ and $B_1$ values for ZPG in figure \ref{fig:fig1} and for pipe flows in figure \ref{fig:fig2}; with corresponding parameter values listed in the captions of all relevant figures in the format ($A_1$, $B_1$). It is noted that in this paper we consider the logarithmic relation for the streamwise normal stresses at face value, as it is predominantly considered in the literature. More recent work by \cite{baars2020part2} and \cite{Desh21} has shown that a pure logarithmic relationship relies on first isolating the very-large-scale or ``superstructure" contributions that do not scale with their distance from the wall. 

The power trend fits, equation \ref{eq:002}, are depicted using dotted lines in figures~\ref{fig:fig1} and \ref{fig:fig2}, with both trends using colors corresponding to the data symbols for each $Re_\tau$. Typical values of $\alpha_1$ and $\beta_1$ as reported in the literature are used. The format in the captions of all relevant figures for the power trend coefficients is ($\beta_1$, $\alpha_1$). In the case of the Superpipe data, the parameters were slightly adjusted to optimize their fit for some of the $Re_\tau$ cases, over the very wide range of Reynolds numbers, and the standard deviation for each parameter is also listed in the caption following the corresponding mean value. The green and blue bars near the bottom of each figure, represent the apparent agreement between the data and the trend models corresponding to logarithmic and power relations, respectively. Two parts of each bar with different shades of its color are used to identify the apparent range of agreement with the corresponding trend for normal-stress profiles with typical lower and higher $Re_\tau$'s.

The thick gray lines are used to provide a visual indication of the boundary between the near-wall and fitting region of the flow, where the validity of the relation will be evaluated more carefully in the following figures.  Utilizing the information in figure 6 of \cite{nag24} and identifying the fitting region as that to the right of the minimum peak in the profiles of turbulent stress gradient divided by viscous stress
gradient, the following relation for the gray lines is used: $y^+ = C_1 (Re_\tau/ C_2)^{0.5}$. This relation recognizes the square root dependence observed in figure 6 of \cite{nag24}. A conservative value of $C_1 = 400$ corresponding to $C_2 = 5,000$ is used. This selects $y^+=400$ as a conservative lower limit of the fitting region for $Re_\tau = 5,000$.

\begin{figure}
      \centering
     \begin{minipage}[t]{0.495\textwidth}
      \centering
      \includegraphics[width=1.0\textwidth]{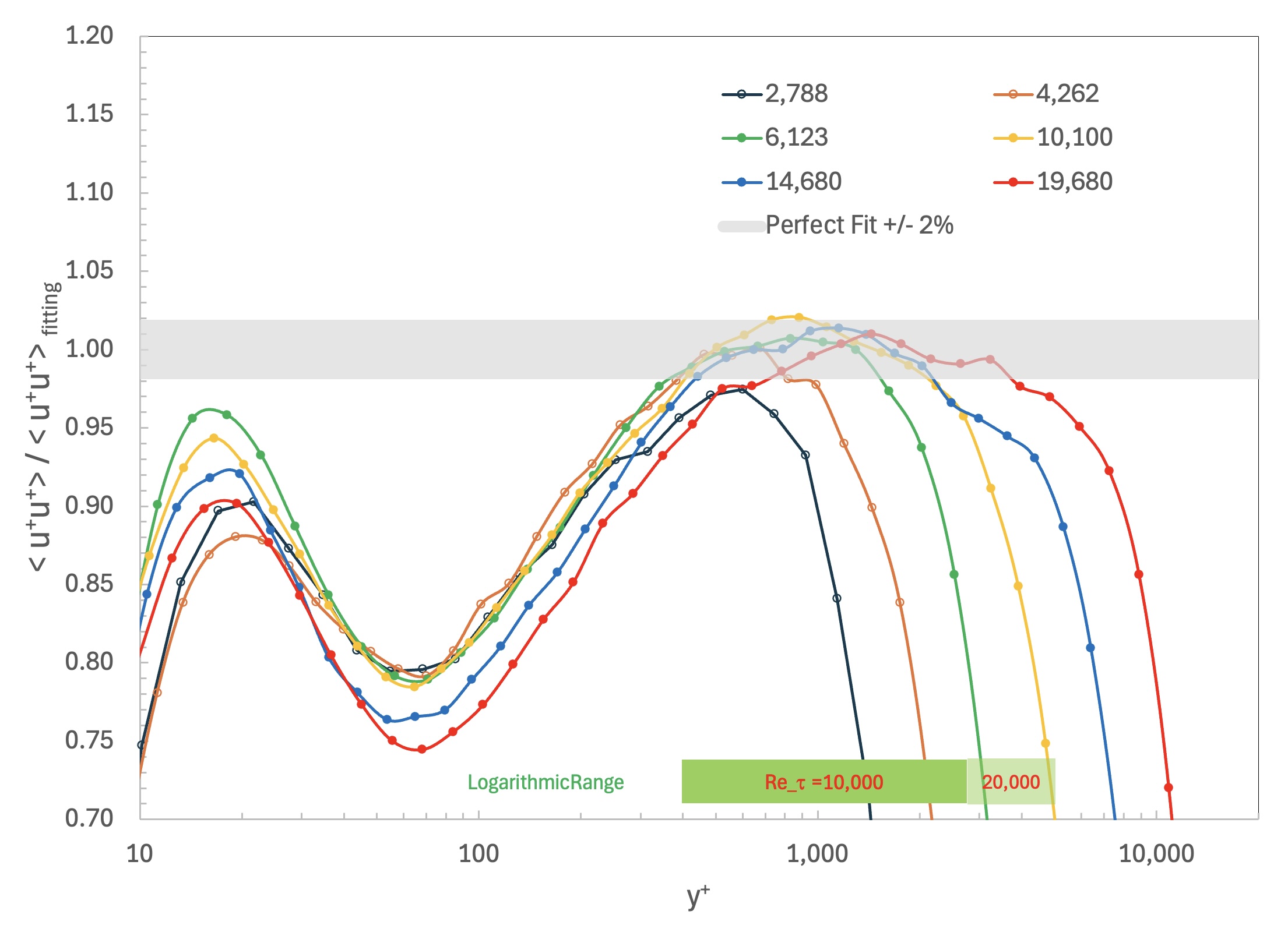}
     \end{minipage}
      \begin{minipage}[t]{0.495\textwidth}
       \centering
      \includegraphics[width=1.0\textwidth]{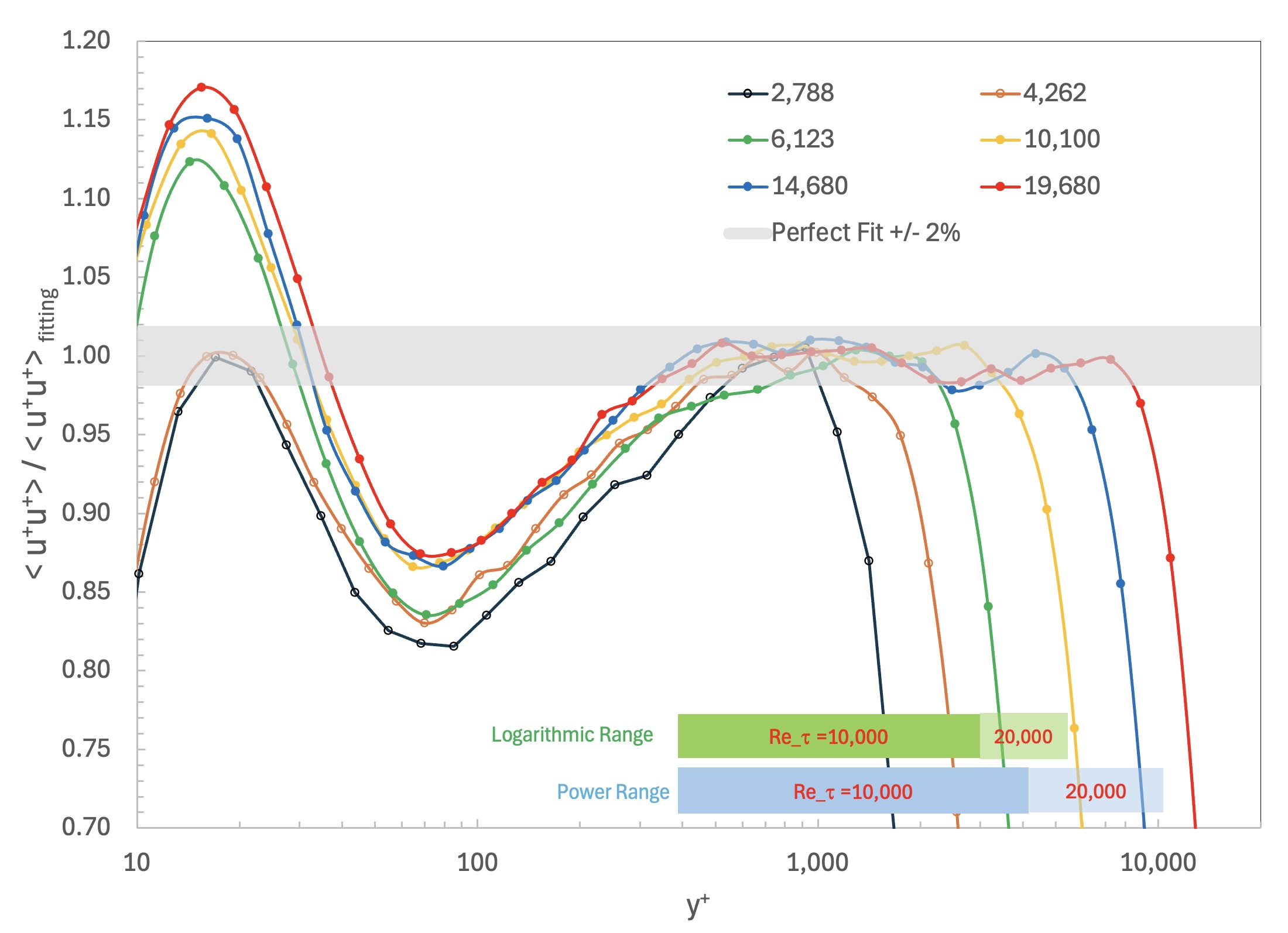}
     \end{minipage}
         \caption{\label{fig:fig3} Streamwise turbulence stress normalized by both fitting relations versus inner-scaled wall distance for ZPG data of \cite{sam18} and \cite{mar15} for different $Re_\tau$'s. (Left) Using \ref{eq:001} with logarithmic parameters (-1.26, 1.93 $\pm$ 0.05). (Right ) Using \ref{eq:002} with 0.25-power parameters (-9.2 $\pm$ 0.22, 10.1 $\pm$ 0.13).}
\end{figure}

\begin{figure}
      \centering
     \begin{minipage}[t]{0.495\textwidth}
      \centering
      \includegraphics[width=1.0\textwidth]{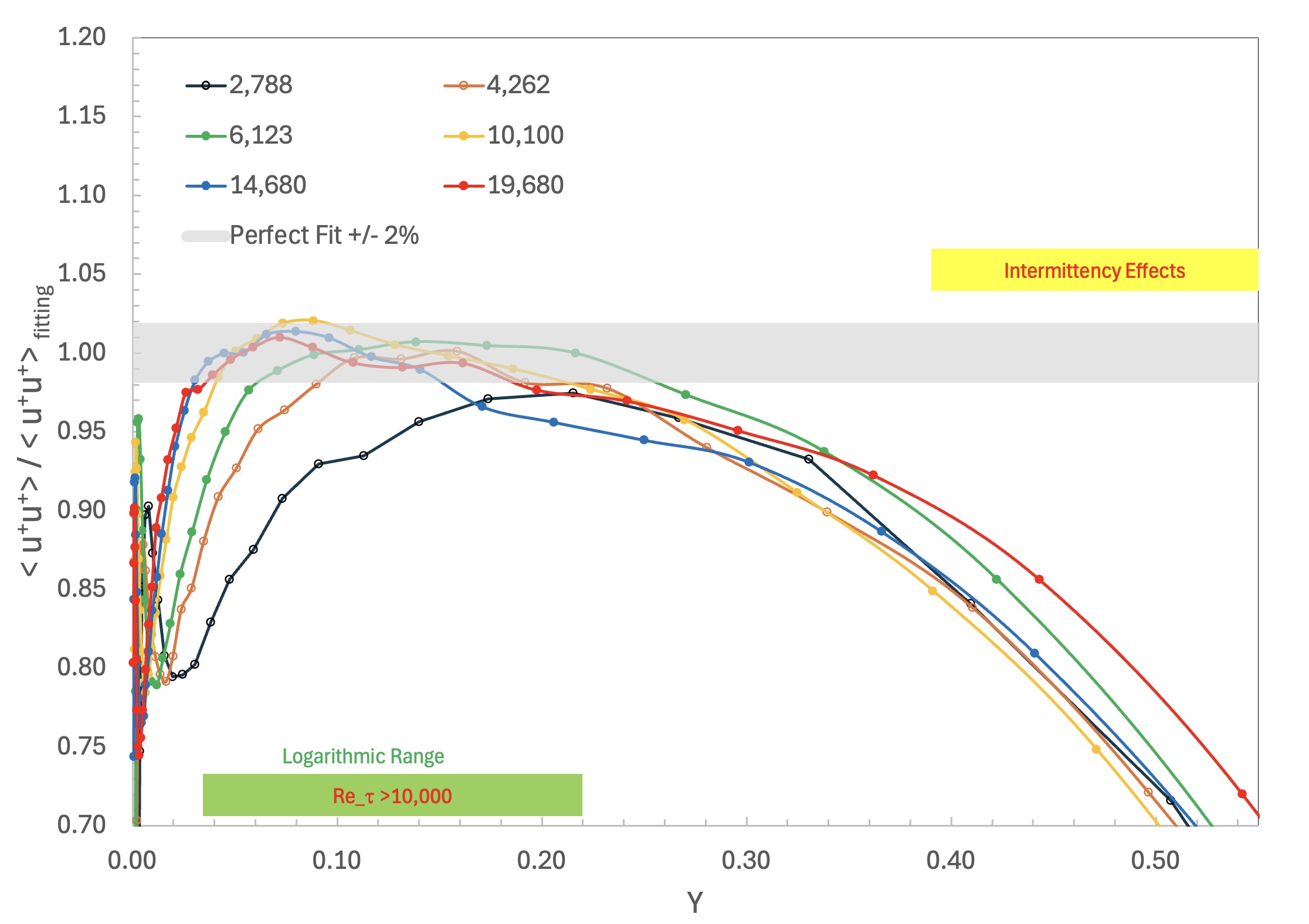}
     \end{minipage}
      \begin{minipage}[t]{0.495\textwidth}
       \centering
      \includegraphics[width=1.0\textwidth]{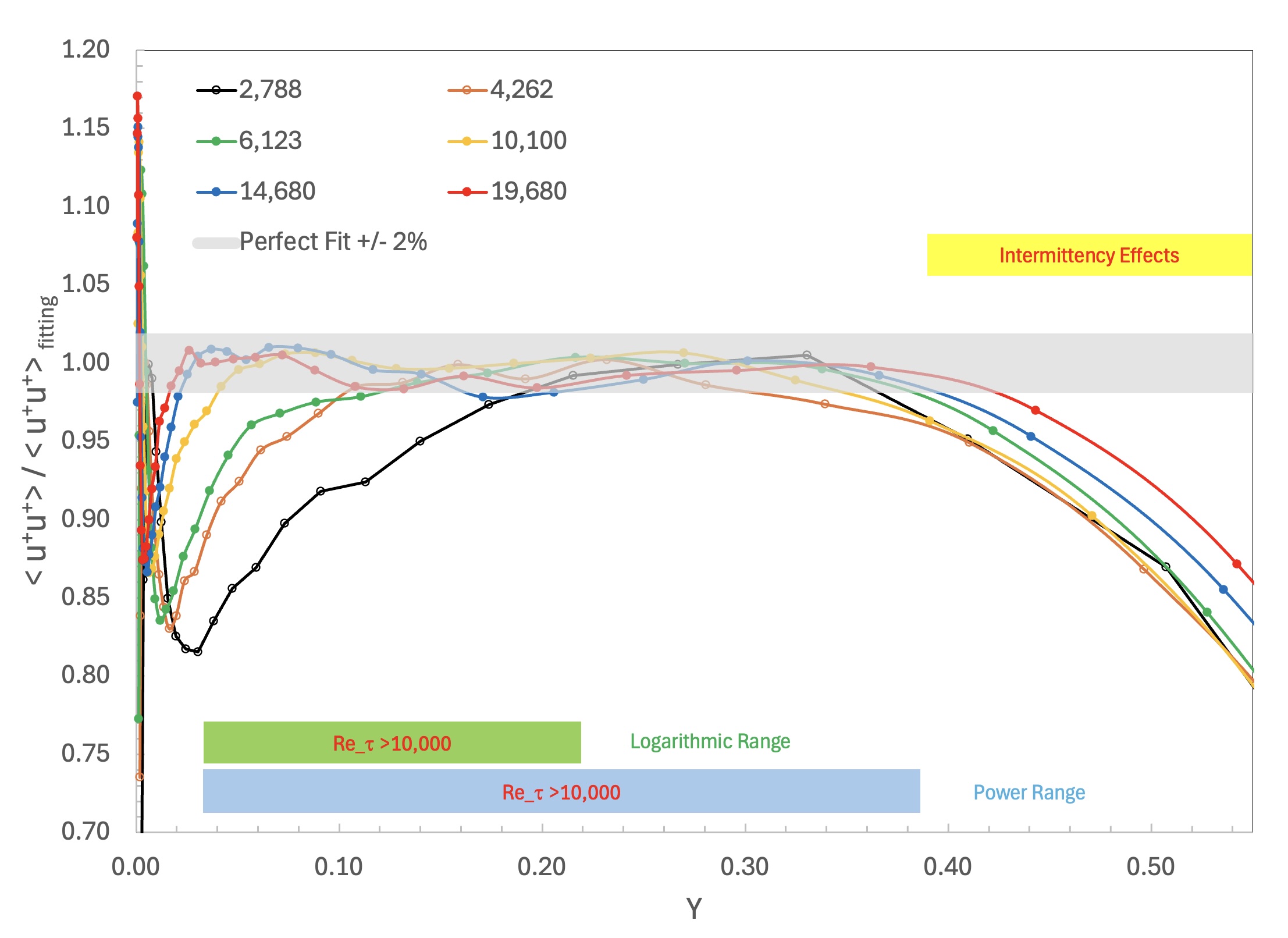}
     \end{minipage}
         \caption{\label{fig:fig4} ZPG data. Same as figure \ref{fig:fig3} but with outer-scaled wall distance.}
\end{figure}

\begin{figure}
      \centering
     \begin{minipage}[t]{0.495\textwidth}
      \centering
      \includegraphics[width=1.0\textwidth]{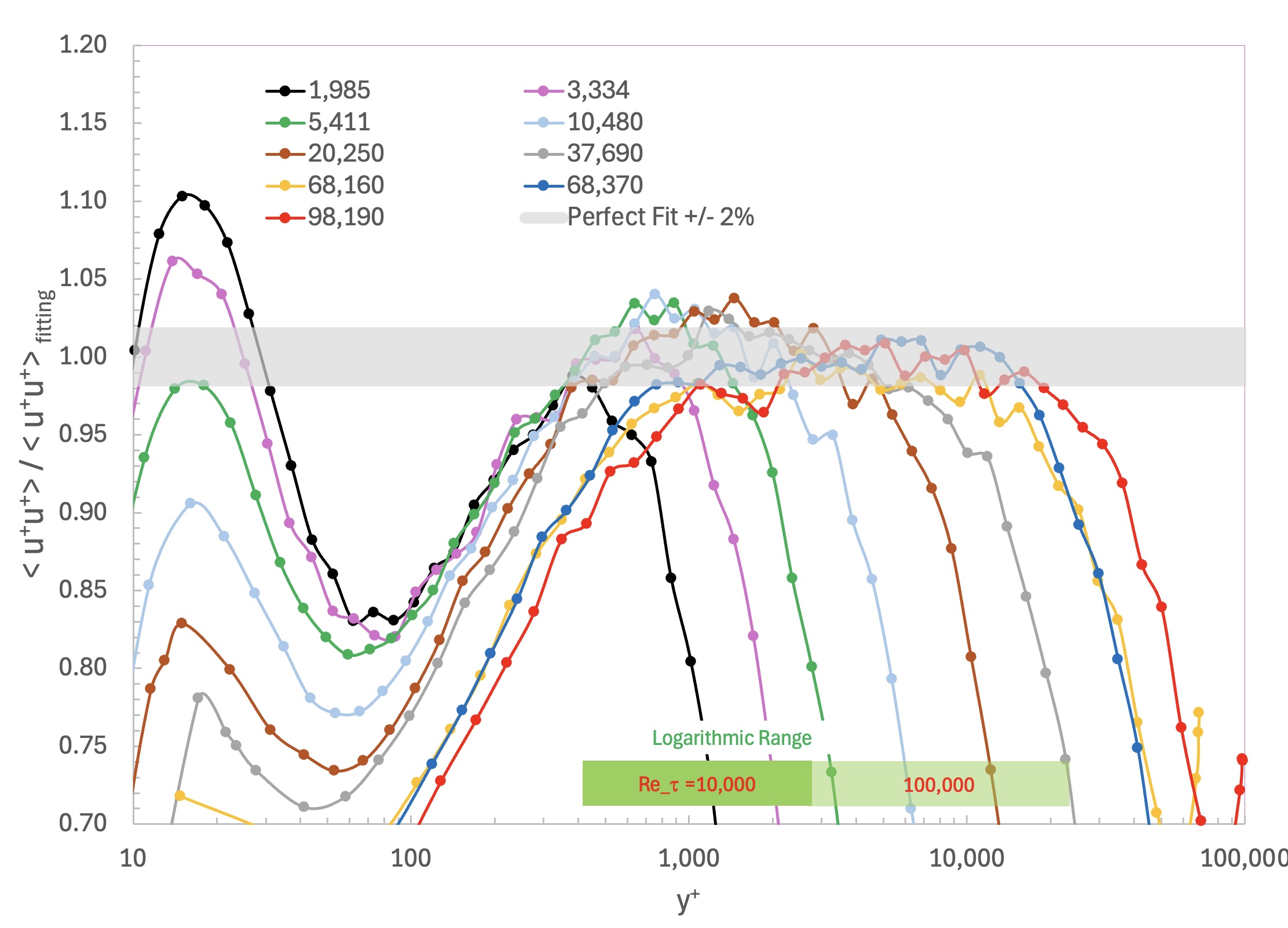}
     \end{minipage}
      \begin{minipage}[t]{0.495\textwidth}
       \centering
      \includegraphics[width=1.0\textwidth]{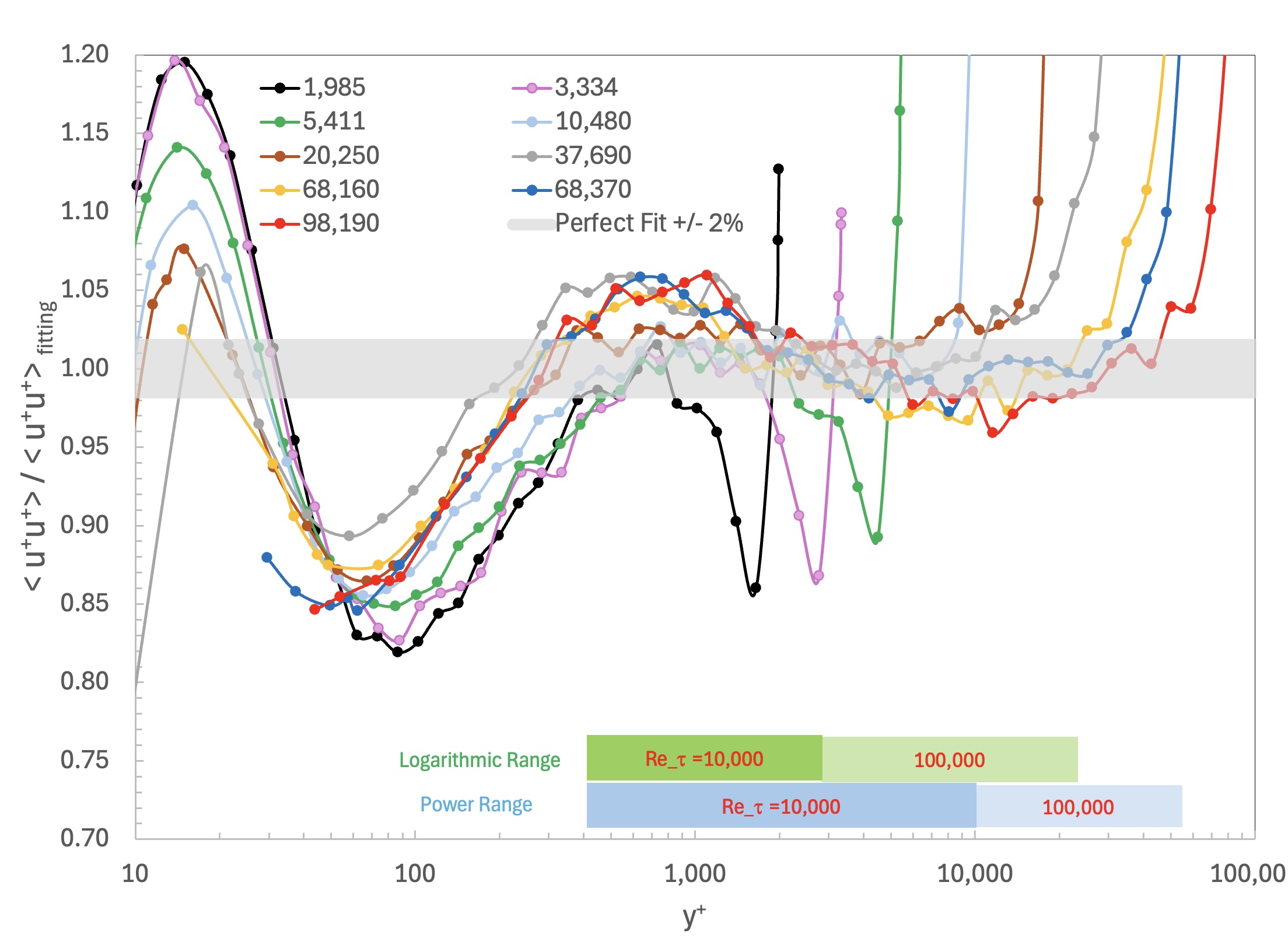}
     \end{minipage}
         \caption{\label{fig:fig5} Streamwise turbulence stress normalized by both fitting relations versus inner-scaled wall distance for Superpipe data of \cite{hul12} for different $Re_\tau$'s. (Left) Using \ref{eq:001} with logarithmic parameters (-1.26, 1.6 $\pm$ 0.17). (Right ) Using \ref{eq:002} with 0.25-power parameters (-9.3 $\pm$ 0.13, 10.0 $\pm$ 0.06).}
     
\end{figure}

\begin{figure}
      \centering
     \begin{minipage}[t]{0.495\textwidth}
      \centering
      \includegraphics[width=1.0\textwidth]{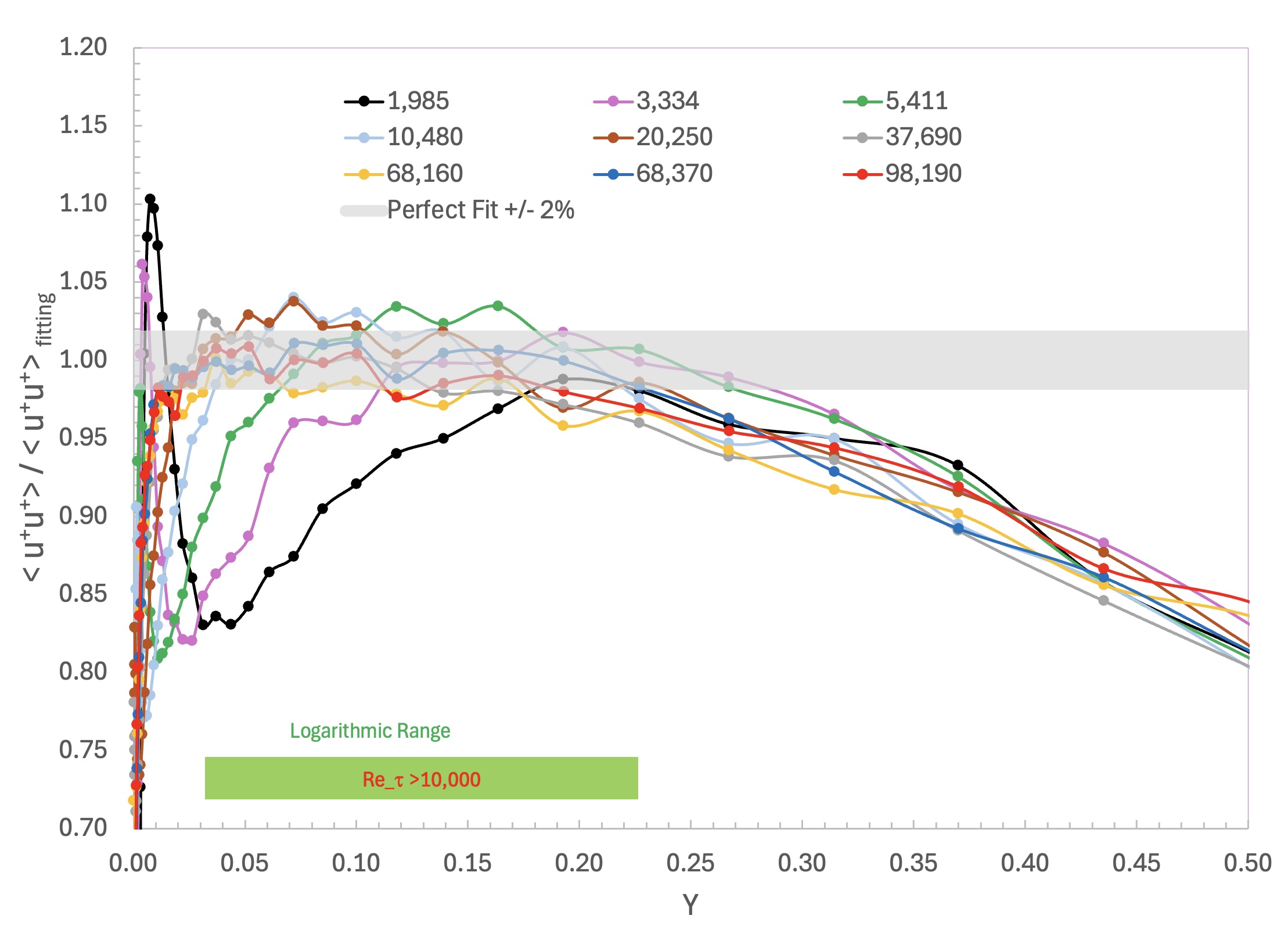}
     \end{minipage}
      \begin{minipage}[t]{0.495\textwidth}
       \centering
      \includegraphics[width=1.0\textwidth]{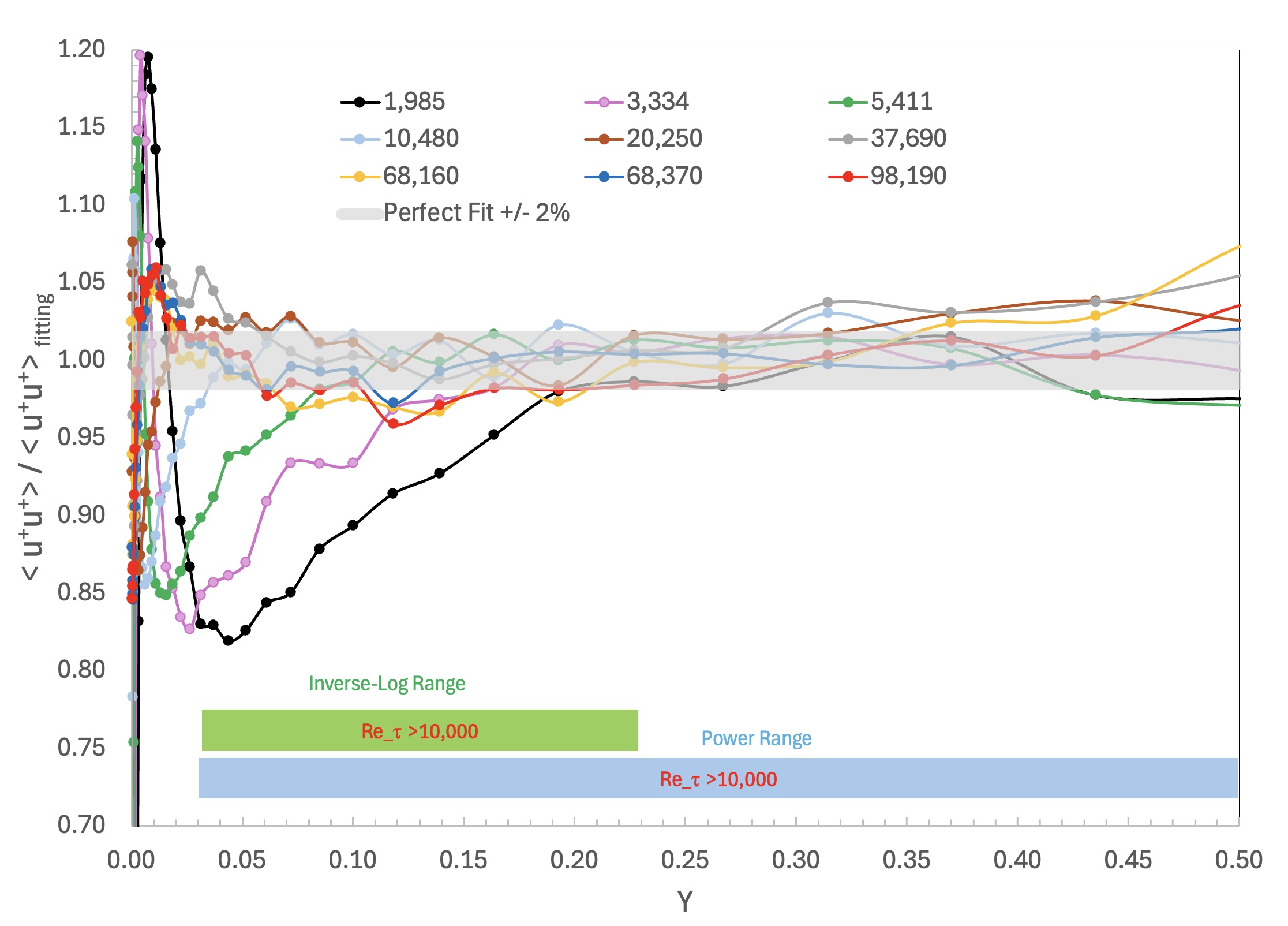}
     \end{minipage}
         \caption{\label{fig:fig6} Superpipe data. Same as figure \ref{fig:fig5} but with outer-scaled wall distance.}
     
\end{figure}

The method introduced here is intended to have a more quantitative evaluation than that used in figures \ref{fig:fig1} and \ref{fig:fig2}, or in most of previous attempts in the literature.  For each model, the normal-stress profile measurements for a given $Re_\tau$ are divided at each data point by the corresponding value from the model and plotted against inner-scaled distance from the wall $y^+$ on a log scale or the outer-scaled distance from the wall $Y$ on a linear scale. To compare the two models, the resulting graphs for the logarithmic model are always placed in the left part of the composite figures \ref{fig:fig3} - \ref{fig:fig6} while the power model is shown in the right side. The values of the parameters for both models $A_1, B_1, \alpha_1$ and $\beta_1$ are listed in the captions for each figure. When various $Re_\tau$ cases required slight adjustments in these parameters, the standard deviation of the full range of $Re_\tau$'s is again listed after the mean value. Based on experience with the entire experimental data sets used here, and recognizing the uncertainties in the measurement of the streamwise normal stress using hot wire sensors, we selected a maximum deviation from the exact agreement ratio of $1.0$ $\pm 2\%$.  In the next figures this range of acceptable model representation of the data is depicted with horizontal gray bands.

Figure \ref{fig:fig3} summarizes the evaluation of the ZPG data using the method described in the previous paragraph. The vertical axis represents the ratio between the measured streamwise normal stress and the expected values based on the fitting trend relation, $\left\langle u^+u^+\right\rangle/\left\langle u^+u^+\right\rangle_{fitting}$. Just as in figures \ref{fig:fig1} and \ref{fig:fig2}, the green and blue bars near the bottom identify the range of validity with the respective fitting trend but more quantitatively now with the help of the horizontal gray bars establishing agreement of the ratio of the data to the relation at the same location of wall distance with $1.0 \pm 0.02$. A similar evaluation is carried out in figure \ref{fig:fig4} using the outer scaled distance $Y$ on a linear horizontal axis. In figures \ref{fig:fig5} and \ref{fig:fig6} the two sets of evaluations versus $y^+$ and $Y$ are repeated for the Superpipe data.  We find from these four figures that the lower limit of fitting of validity in inner variables for both logarithmic and power relations  changes weakly  for $Re_\tau \gtrapprox 10,000$ within the scatter of the results. 

Examining the minimum peak in both parts of Figure 6 of \cite{nag24} and relating them to an arbitrary fractional power of $Re_\tau$ with an adjustable proportionality coefficient, we find that a best fit is $y^+ = 1.5 \cdot Re_\tau^{0.5}$. In the figure, this minimum peak is readily identified with a transition between the near wall flow and the overlap region over the range $1,000 < Re_\tau < 16,000$ based on DNS of channel flow and, therefore, readily with $y^+_{in}$.

Consistent with the analysis of \citet{Klew09} and \citet{nag24} we will thus consider $y^+_{in} \sim (Re_\tau)^{0.5}$, with the provision that $y^+_{in} \gtrapprox 400$ for the ZPG and Superpipe data. This corresponds to $0.004 \lessapprox Y_{in} \lessapprox 0.04$ for the range of $Re_\tau$ considered here. The upper limit of fitting validity in outer variables $Y_{out}$, for both logarithmic and power relations, appears to be independent of $Re_\tau$ in the Superpipe and ZPG data; see figures \ref{fig:fig4},  \ref{fig:fig6} and \ref{fig:fig7}.  We find bounds for the logarithmic relation  that incorporate the region of the flow where the classical pure-logarithmic relation of the mean velocity profiles can be identified (see region $Y \lessapprox 0.1$ in Figure 4 of \cite{bax24} and Figure 7 of \cite{mon23}), and therefore, consistent with the wall-scaled (attached-) eddy model \citep{mar13}. However, it is significant to note that the upper limit of fitting validity in outer variable $Y_{out}$ is considerably higher for the power relation compared to the logarithmic relation. For ZPG, $Y_{out}$ is $0.39$ compared to $0.22$, and for the Superpipe data is $0.5$ versus $0.27$; see figures \ref{fig:fig4} and \ref{fig:fig6}. The corresponding limits in $y^+$ values are easily obtained from the $Re_\tau$ conditions. For example, when $Re_\tau \approx 20,000$ the logarithmic trend is found only up to $y^+\approx 5,000$ while the power relation extends the agreement to $y^+\approx 9,000$.

\section{Power-relation exponent, intermittency effects, and comparison to DNS}

\begin{figure}
      \centering
     \begin{minipage}[t]{0.495\textwidth}
      \centering
      \includegraphics[width=1.0\textwidth]{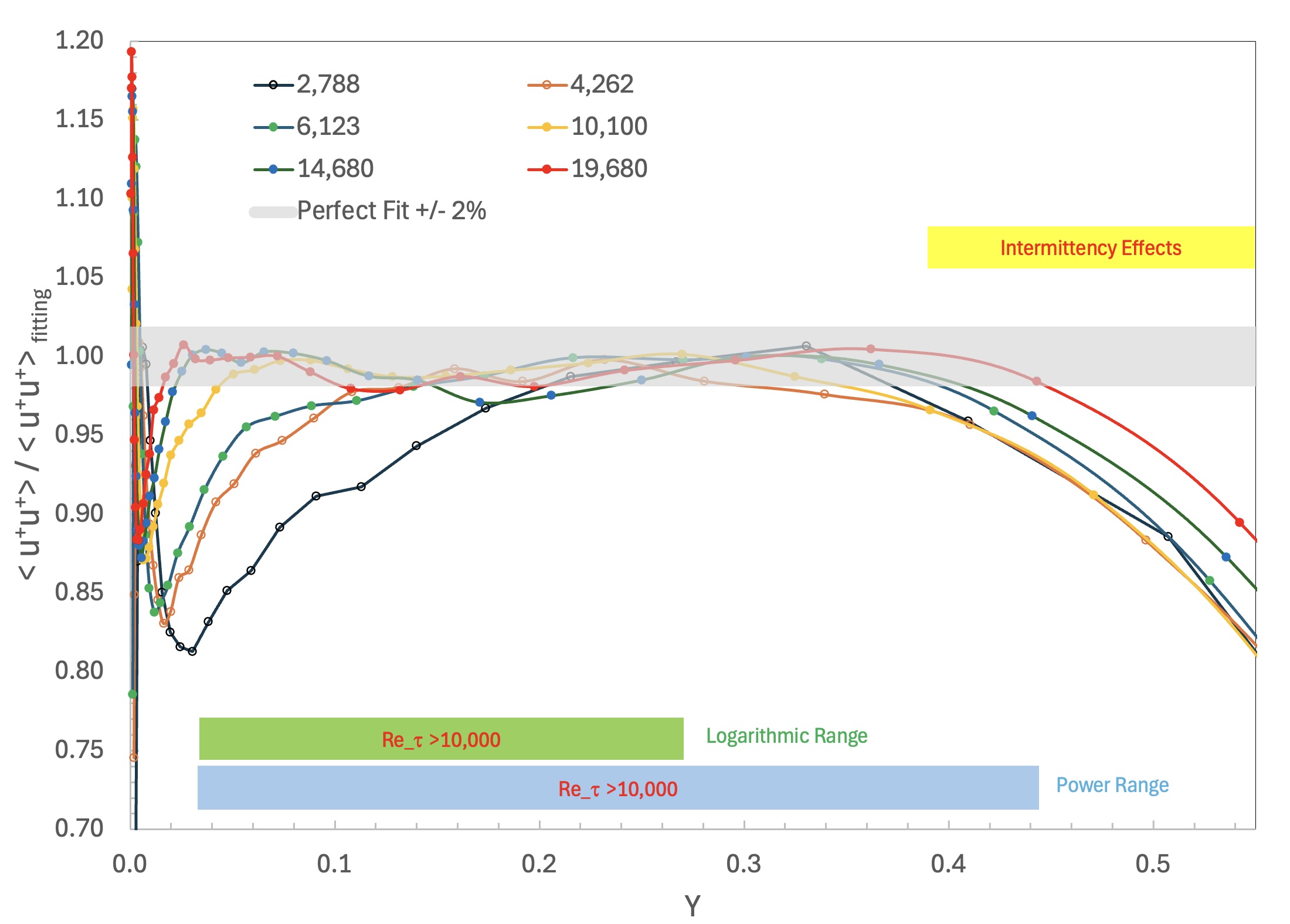}
     \end{minipage}
      \begin{minipage}[t]{0.495\textwidth}
       \centering
      \includegraphics[width=1.0\textwidth]{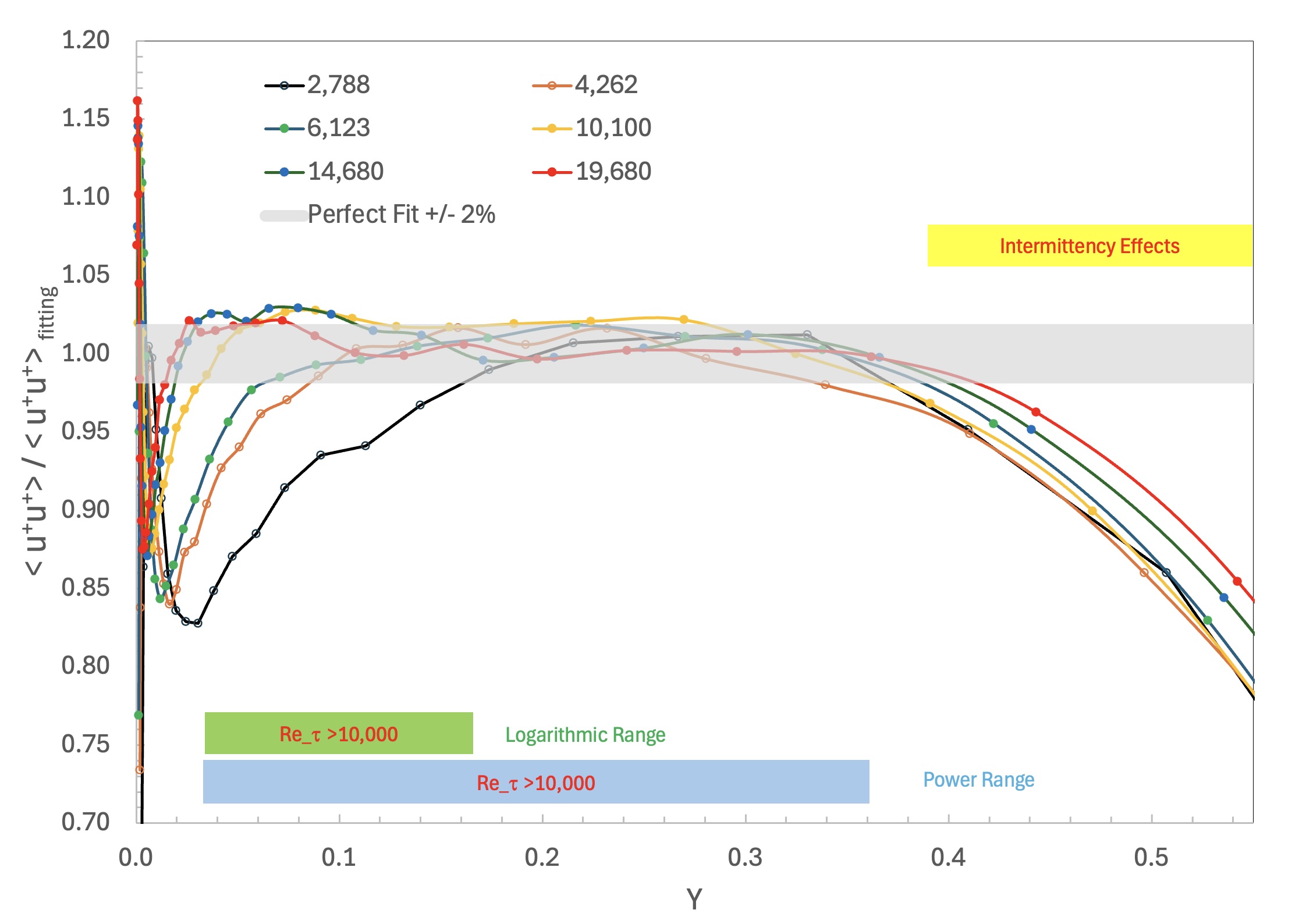}
     \end{minipage}
    \caption{\label{fig:fig7} Streamwise turbulence stress normalized by both fitting relations versus outer-scaled wall distance for different $Re_\tau$'s. (Left) \cite{sam18} ZPG data with
0.28-power parameters (-8.9 $\pm$ 0.21, 9.6 $\pm$ 0.26). (Right) \cite{sam18} ZPG data with
0.22-power parameters (-9.6 $\pm$ 0.23, 10.6 $\pm$ 0.13).}
     
\end{figure}

\begin{figure}
      \centering
     \begin{minipage}[t]{0.495\textwidth}
      \centering
      \includegraphics[width=1.0\textwidth]{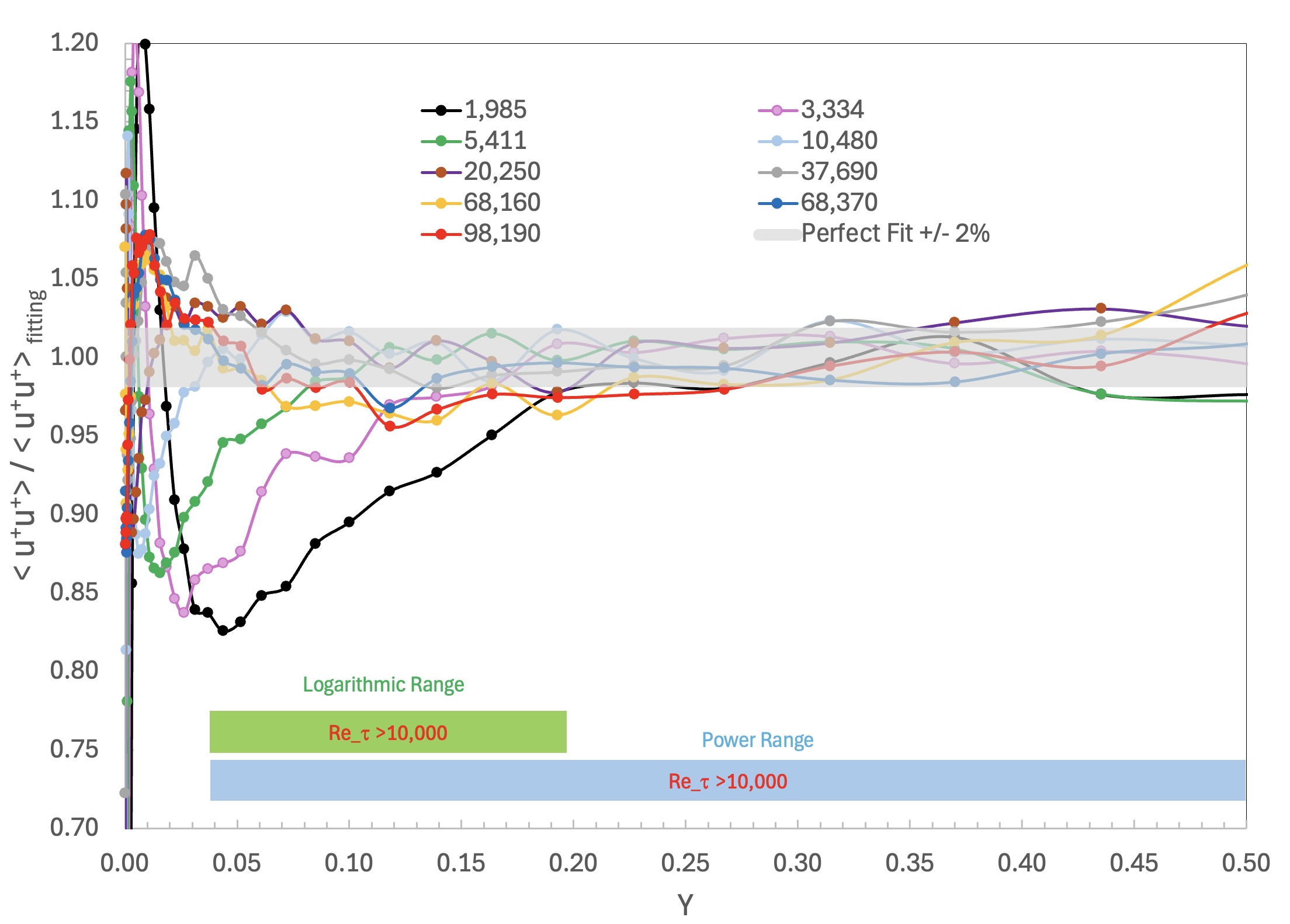}
     \end{minipage}
      \begin{minipage}[t]{0.495\textwidth}
       \centering
      \includegraphics[width=1.0\textwidth]{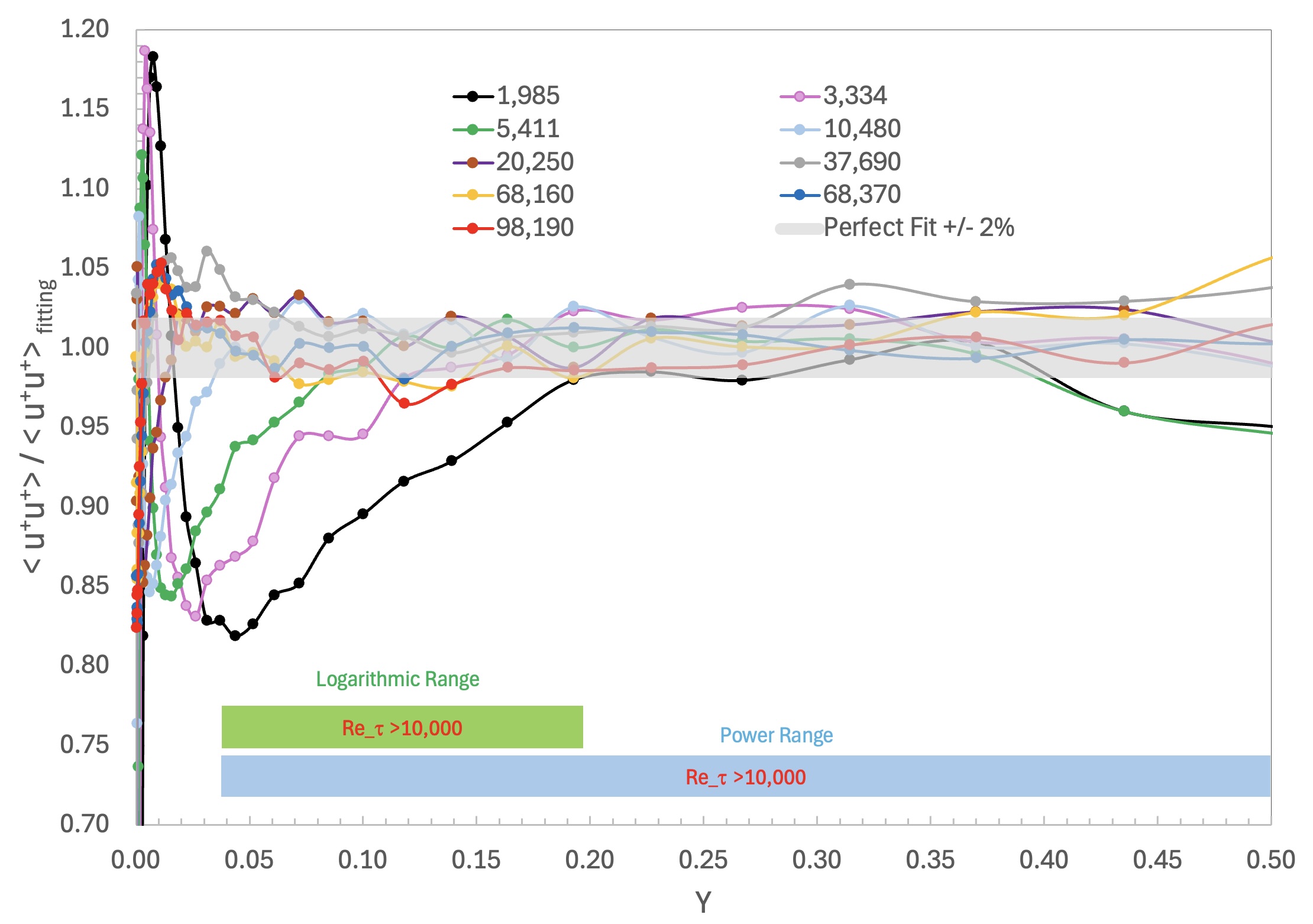}
     \end{minipage}
    \caption{\label{fig:fig8} Streamwise turbulence stress normalized by both fitting relations versus outer-scaled wall distance for different $Re_\tau$'s. (Left) \cite{hul12} Superpipe data with
0.28-power parameters (-8.6 $\pm$ 0.1, 9.10 $\pm$ 0.23). (Right) \cite{hul12} Superpipe data with
0.22-power parameters (-9.8 $\pm$ 0.1, 10.5 $\pm$ 0.26).}
     
\end{figure}

\begin{figure}
      \centering
     \begin{minipage}[t]{0.495\textwidth}
      \centering
      \includegraphics[width=1.0\textwidth]{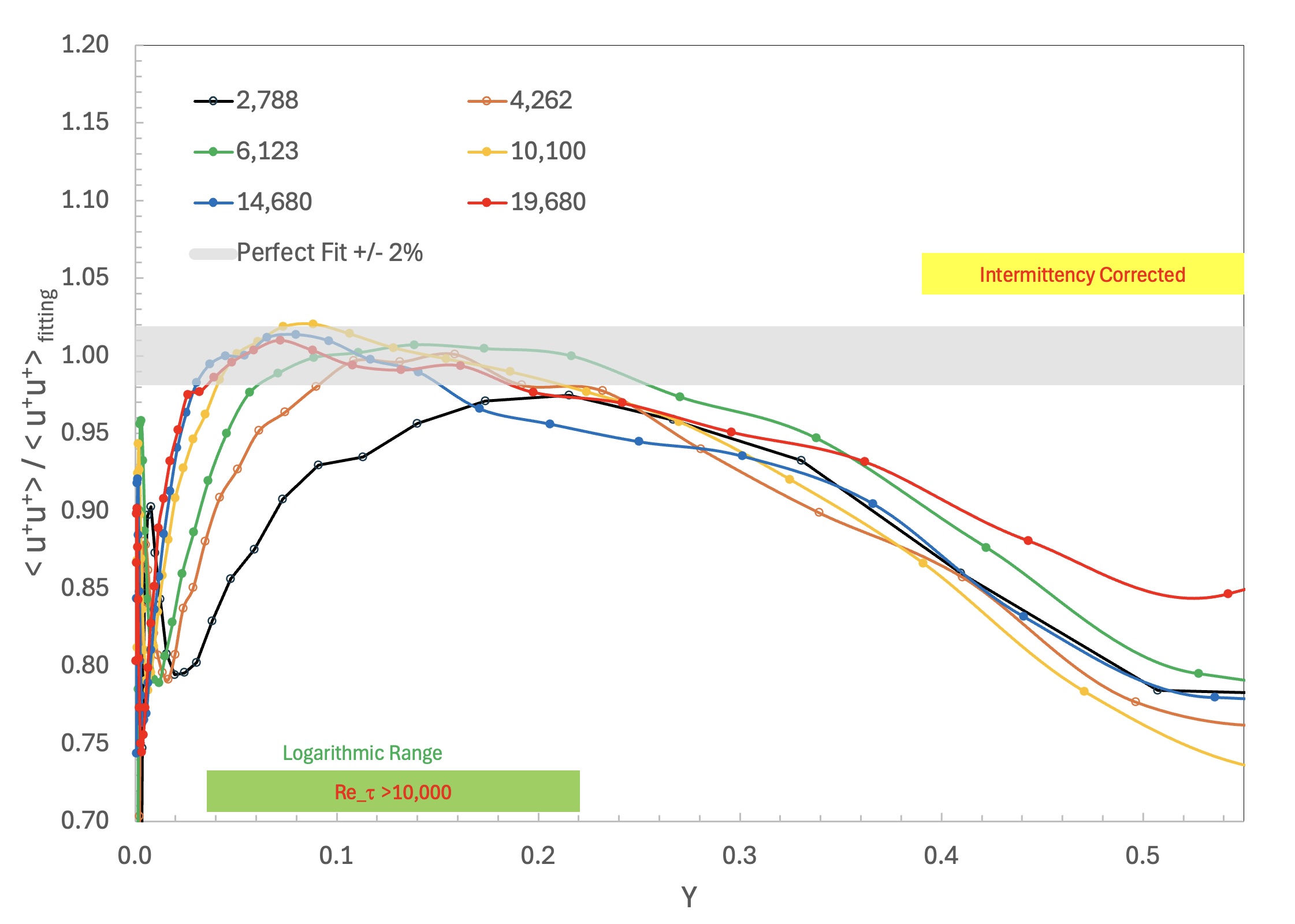}
     \end{minipage}
      \begin{minipage}[t]{0.495\textwidth}
       \centering
      \includegraphics[width=1.0\textwidth]{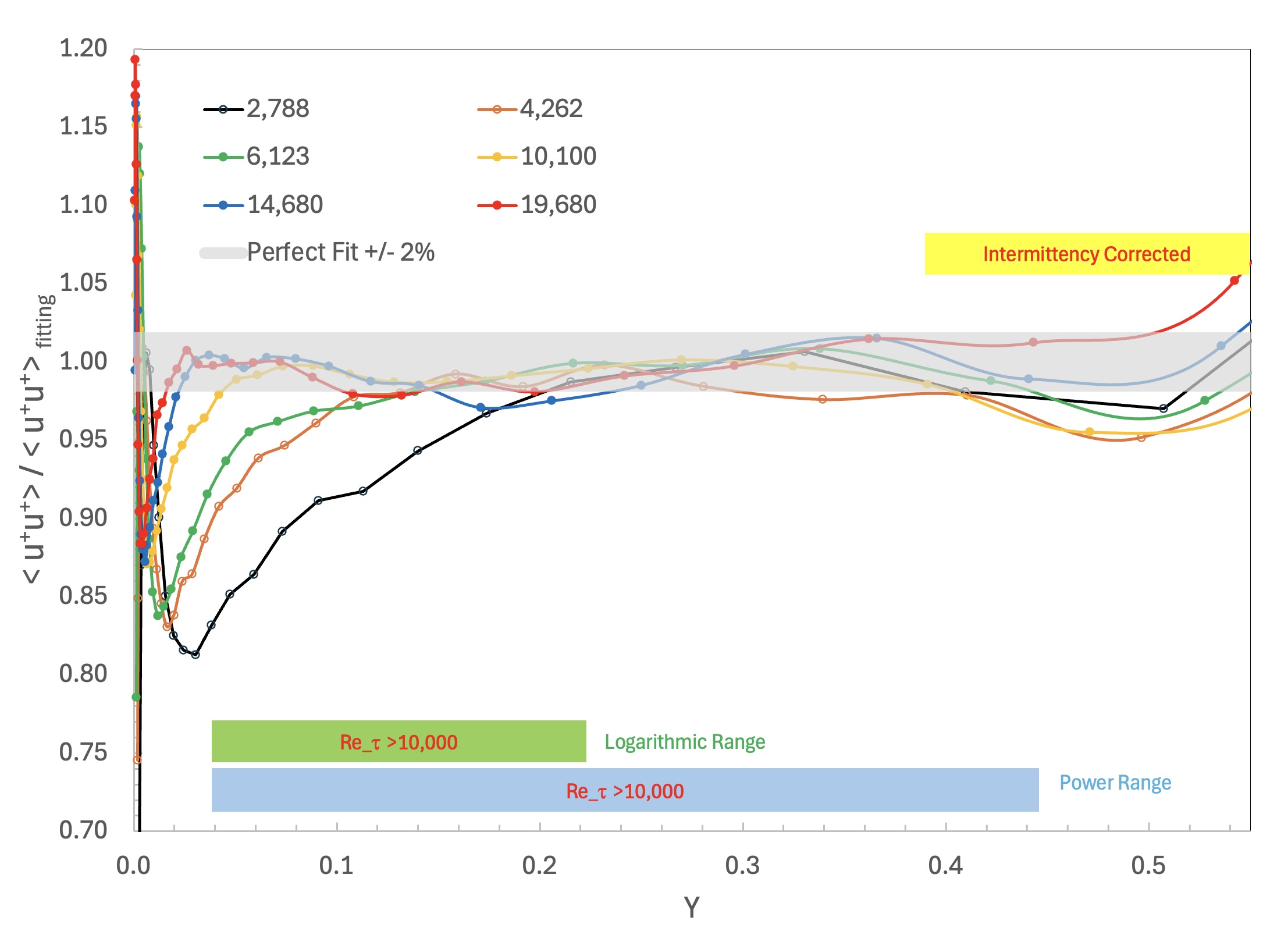}
     \end{minipage}
    \caption{\label{fig:fig9} Streamwise turbulence stress normalized by both fitting relations versus outer-scaled wall distance for different $Re_\tau$'s, with intermittency correction applied. (Left) \cite{sam18} and \cite{mar15} ZPG data (open symbols) with logarithmic  parameters (-1.26, 1.93 $\pm$ 0.05).
 (Right) \cite{sam18} and \cite{mar15} ZPG data (open symbols) with
0.28-power parameters (-8.9 $\pm$ 0.21, 9.6 $\pm$ 0.16). }
     
\end{figure}

In figures \ref{fig:fig7} and \ref{fig:fig8},  the variation of the exponent of the power model is tested. The left side of each figure displays the evaluation of the experimental data by a power relation with $0.28$ exponent, while the right side uses an exponent of $0.22$. The ZPG data is evaluated in figure \ref{fig:fig7} and the range of potential impact of the outer intermittency is indicated for reference by a yellow bar near the top of the figure. In contrast, for the Superpipe data shown in figure \ref{fig:fig8}, no influence of intermittency is expected and the agreement with the power relation extends to half the pipe radius.  Comparing the results from both ZPG and Superpipe data for power relations with exponents of $0.22$, $0.25$ and $0.28$, reveals that a exponent of $0.28$ represents the data somewhat better than the original $0.25$-power relation based on the bounded dissipation model; see also figures \ref{fig:fig11} and \ref{fig:fig12}. Several other exponents were tested with both data sets, but the $0.28$ appears to yield the best agreement. Applying the intermittency correction used for figure \ref{fig:fig9} to the $0.28$-power relation yields the best agreement with the measured data including for large $y^+$ values. The fitting range agreement with the $0.28$-power model and the intermittency correction extends to $Y$ over $0.5$, consistent with the result of the ratio method in figure \ref{fig:fig9}.

Correcting ZPG data for outer intermittency effects using the function of \cite{Chau14}, does not change $Y_{out}$ for the logarithmic relation, while extending $Y_{out}$ for the power relation to around $0.44$; see figure \ref{fig:fig9}.  The correction is simply applied by dividing the ratio $\left\langle u^+u^+\right\rangle/\left\langle u^+u^+\right\rangle_{fitting}$ by the intermittency function that ranges from one at the wall to zero beyond $y = \delta$.

\begin{figure}
      \centering
     \begin{minipage}[t]{0.495\textwidth}
      \centering
      \includegraphics[width=1.0\textwidth]{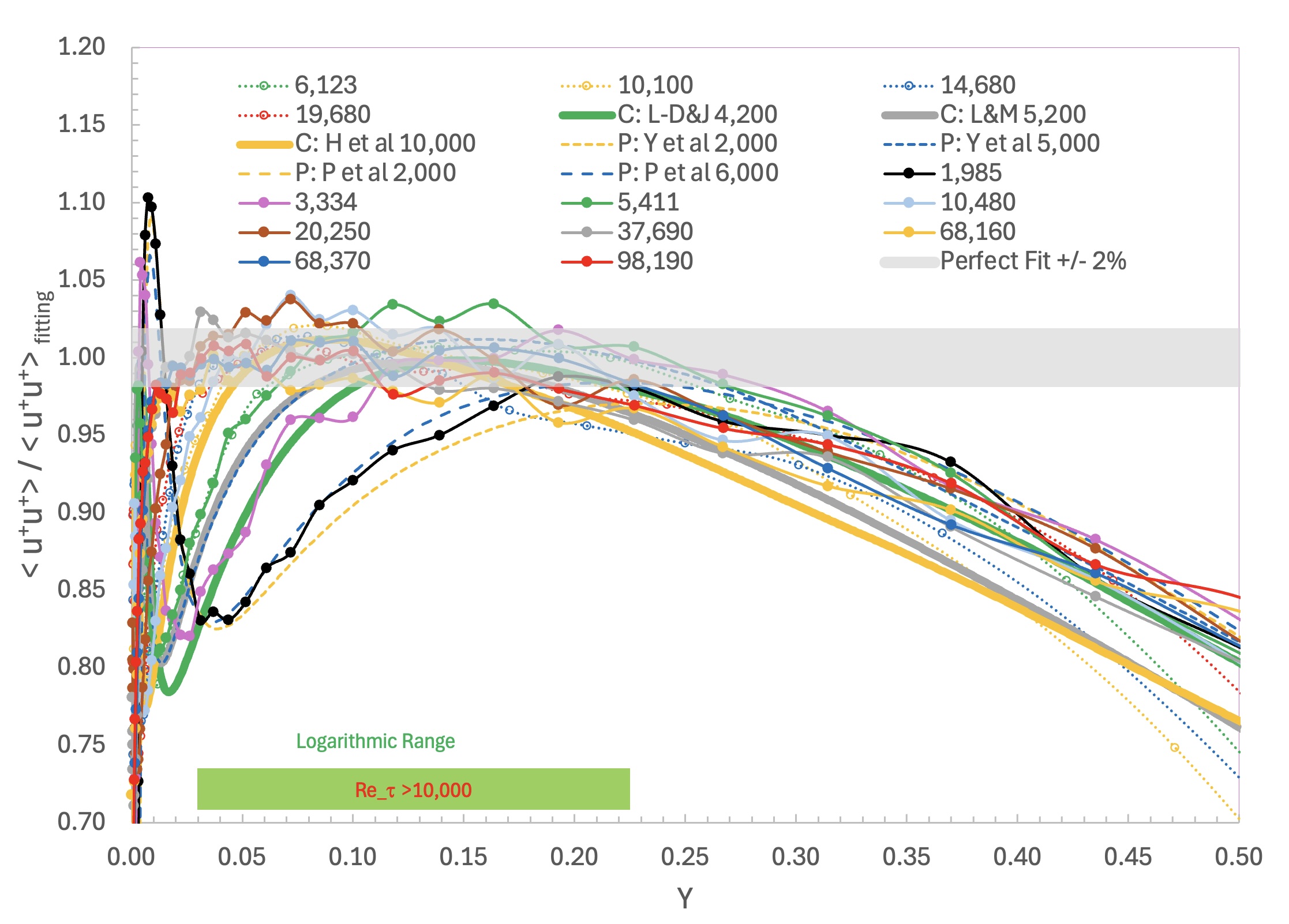}
     \end{minipage}
      \begin{minipage}[t]{0.495\textwidth}
       \centering
      \includegraphics[width=1.0\textwidth]{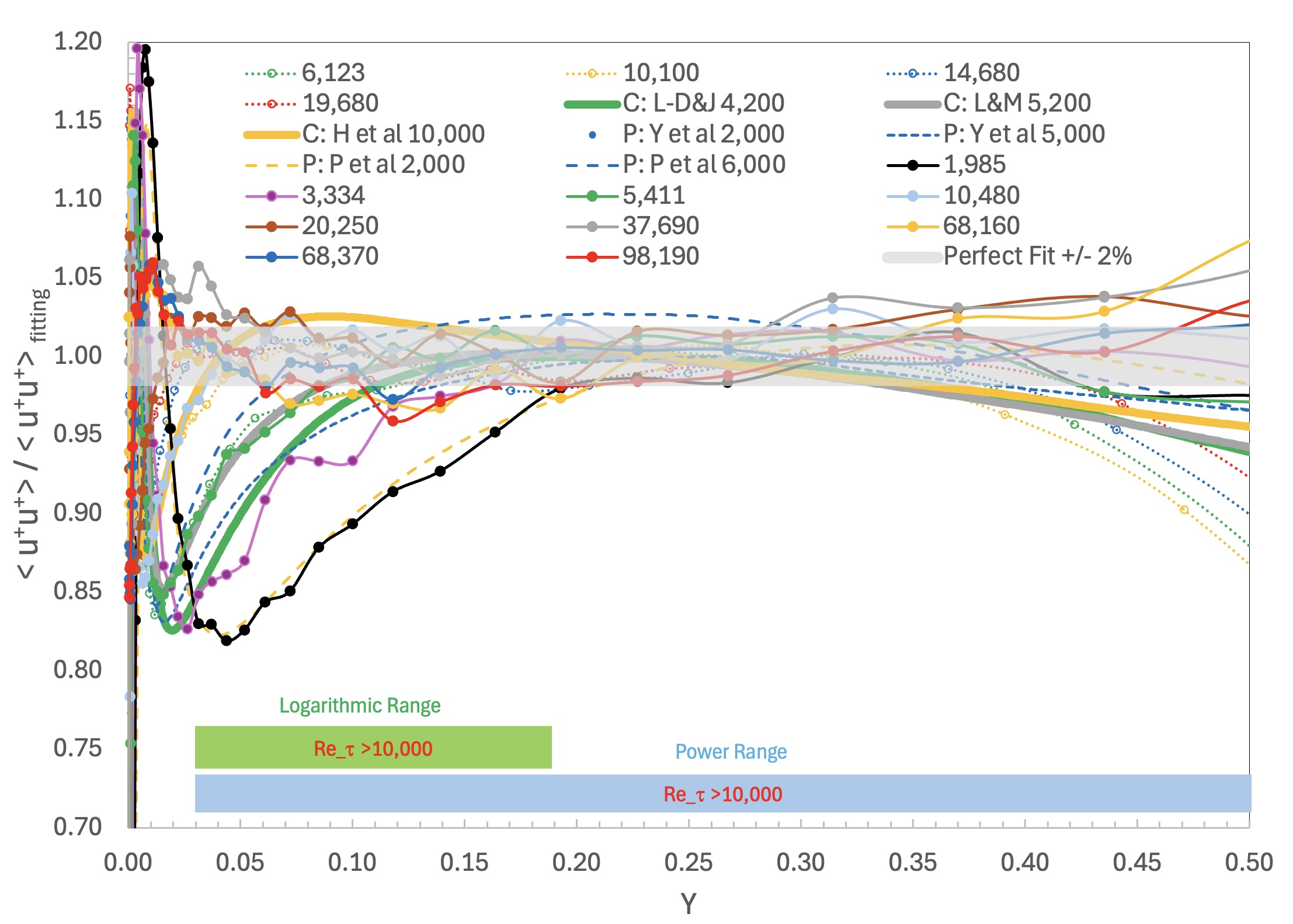}
     \end{minipage}
    \caption{\label{fig:fig10} Streamwise turbulence stress normalized by both fitting relations versus outer-scaled wall distance for different $Re_\tau$'s. (Left) \cite{hul12} Superpipe data with
logarithmic parameters (-1.26, 1.6 $\pm$ 0.17),
\cite{sam18} ZPG (open symbols and dotted lines) and Channel DNS data from \cite{nag24}. (Right) \cite{hul12} Superpipe data with
0.25-power parameters (-9.3 $\pm$ 0.13, 10.0 $\pm$ 0.06),
\cite{sam18} ZPG and Channel DNS data from \cite{nag24}.}
     
\end{figure}

\begin{figure}
       \centering
      \includegraphics[width=0.85\textwidth]{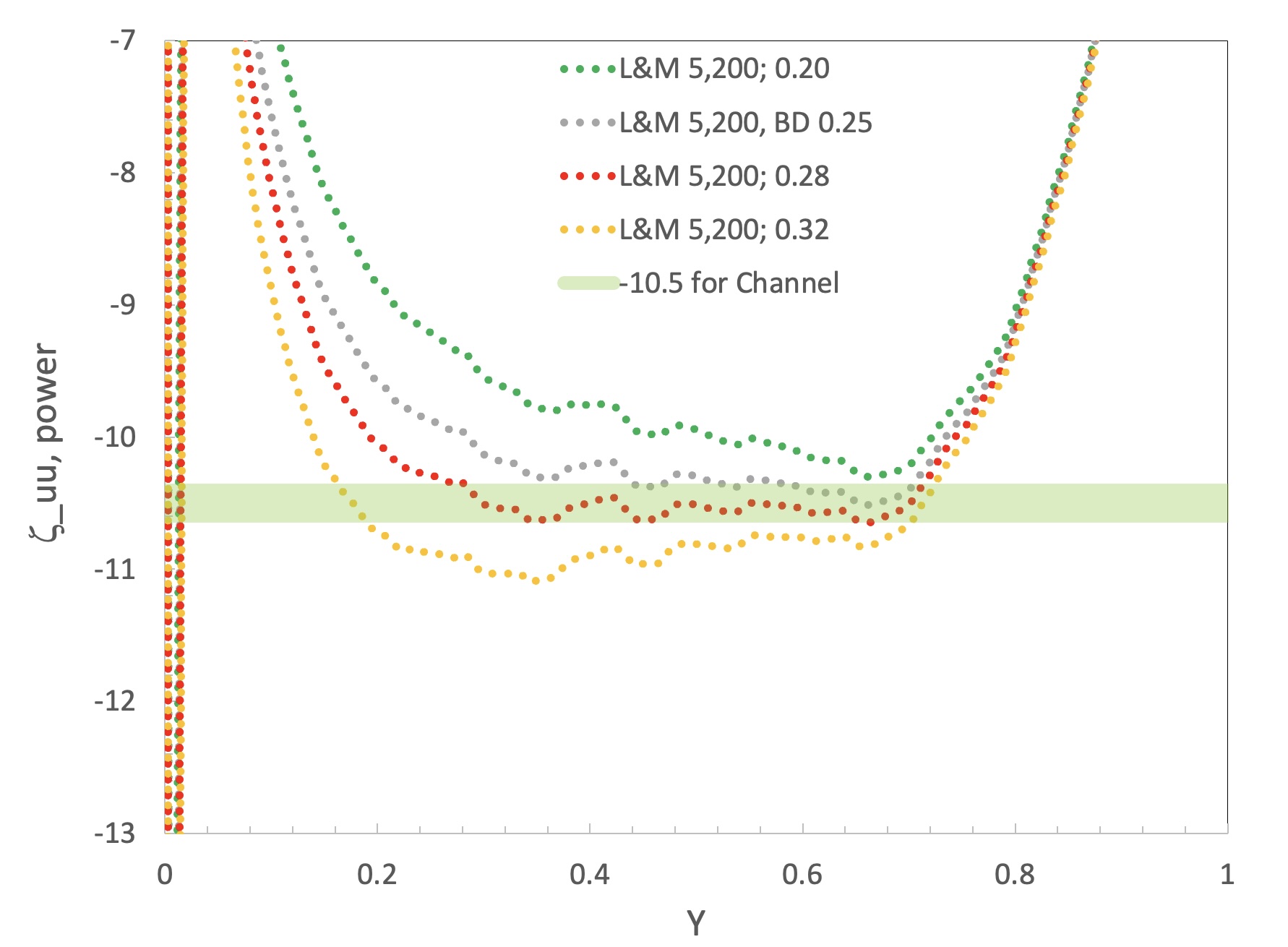}
    \caption{\label{fig:fig11} Indicator functions of  normal stress computed from DNS results of \cite{lee15} for power relations with exponent varying between $0.2$ and $0.32$.}
     
\end{figure}

\begin{figure}
      \centering
      \includegraphics[width=0.95\textwidth]{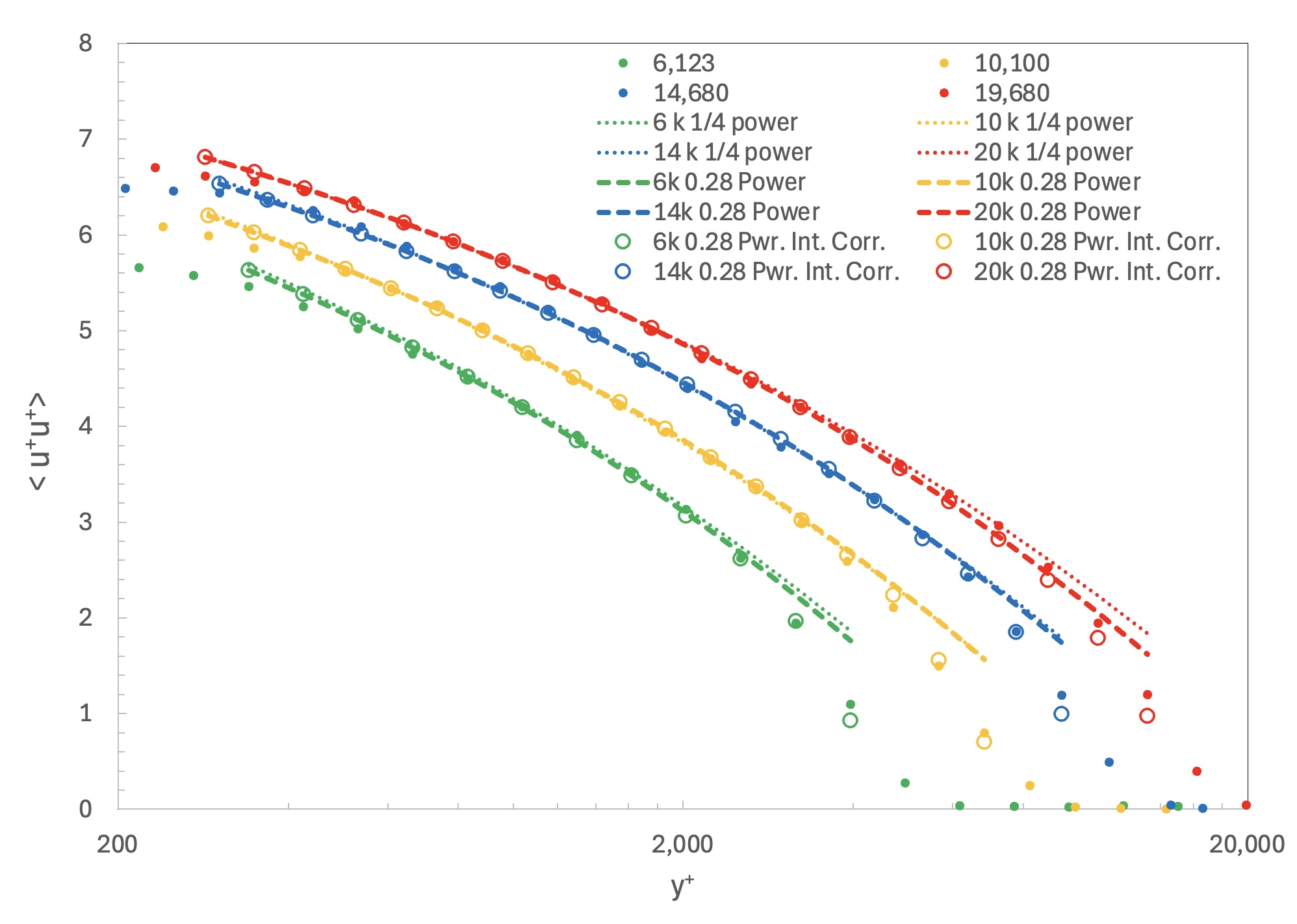}
    \caption{\label{fig:fig12} Streamwise normal stress versus $Re_\tau$ for ZPG data of \cite{sam18}, comparing trends of 0.25-power with parameters (-9.2 $\pm$ 0.22, 10.1 $\pm$ 0.13), 0.28-power with parameters (-8.9 $\pm$ 0.21, 9.6 $\pm$ 0.26), and intermittency correction of 0.28-power trend.}
\end{figure}

In figure \ref{fig:fig10}, DNS results of pipe flow are displayed with dashed lines and channel flow data with continuous lines. Dashed lines with experimental data points identified by solid circles are used for Superpipe data and continuous lines with experimental data points marked by solid circles are used for ZPG data.  The color pattern for the experimental data for different $Re_\tau$ conditions are the same as in earlier figures.

Comparing the experimental data for ZPG and Superpipe to DNS data for pipe flow in the range $2,000 < Re_\tau < 6,000$ and channel flow over $2,000 < Re_\tau < 10,000$, reveals general agreement for results of both logarithmic and power relations as displayed in figure \ref{fig:fig10}. The DNS results are based on the recent work of \cite{nag24}, and all the parameters of the various cases are given in tables included by them.  

Figure \ref{fig:fig10} is a further demonstration that, especially near the wall, $Re_\tau \gtrapprox 10,000$ is required for the two proposed relations to achieve agreement with the streamwise normal stress in the fitting region. This is in contrast to a lower value of $Re_\tau \gtrapprox 5,000$ we find for the indicator functions of the mean velocity profile, as demonstrated by \cite{hoy24b} and figure 1 of \cite{nag24}.

Next, we utilize the $Re_\tau = 5,200$ DNS data of \cite{lee15} for channel flow, and the indicator function for the power trend given by \cite{nag24} for $0.25$ power given by:

\begin{equation}\label{eq:003}
 \zeta_{uu,BD} = 4Re_\tau^{0.25}(y^+)^{0.75}\frac{{\rm d}{\left\langle u^+u^+\right\rangle}}{{\rm d}y^+} = 4Y^{0.75}\frac{{\rm d}{\left\langle u^+u^+\right\rangle}}{{\rm d}Y}.   
\end{equation}

 To evaluate the different values of the exponent using such an indicator function we use:
\begin{equation}\label{eq:004}
 \zeta_{uu,power} = F_{power} \cdot \zeta_{uu,BD},
\end{equation}
and select $F_{power}$ for each exponent to bring the indicator functions close to the expected value for channel flow of $-10.5$, in order to achieve an optimum visual comparison with increasing value of exponent. The best agreement, with widest range in $Y$, found with the data for several values of the power exponent is clearly the exponent of $0.28$, for which we select $F_{0.28}$ to match the $-10.5$ value for $\zeta_{uu,power}$. The results are summarized in figure \ref{fig:fig11} for four different exponents of the power relation. They reveal that a value of $0.28$ for the exponent, instead of the $0.25$ value predicted by \cite{che22, che23},  produces the widest range in wall distance with agreement.  This result is also supported by the data of \cite{sam18} for ZPG boundary layers as demonstrated by figure \ref{fig:fig13}, where the dashed lines are closer to the data especially at large $y^+$ values.

\section{Projecting peak streamwise normal stress}

Here, we consider possible implications of the trends with $Re_\tau$ of the two relations, logarithmic and power, for the inner peak of the streamwise normal stress, which is usually found around $y^+ \approx 15$. This will be done by projecting the values of the normal stress in the very near wall region from the fitting region, with $y^+ > 400$.

\begin{figure}
      \centering
      \includegraphics[width=0.95\textwidth]{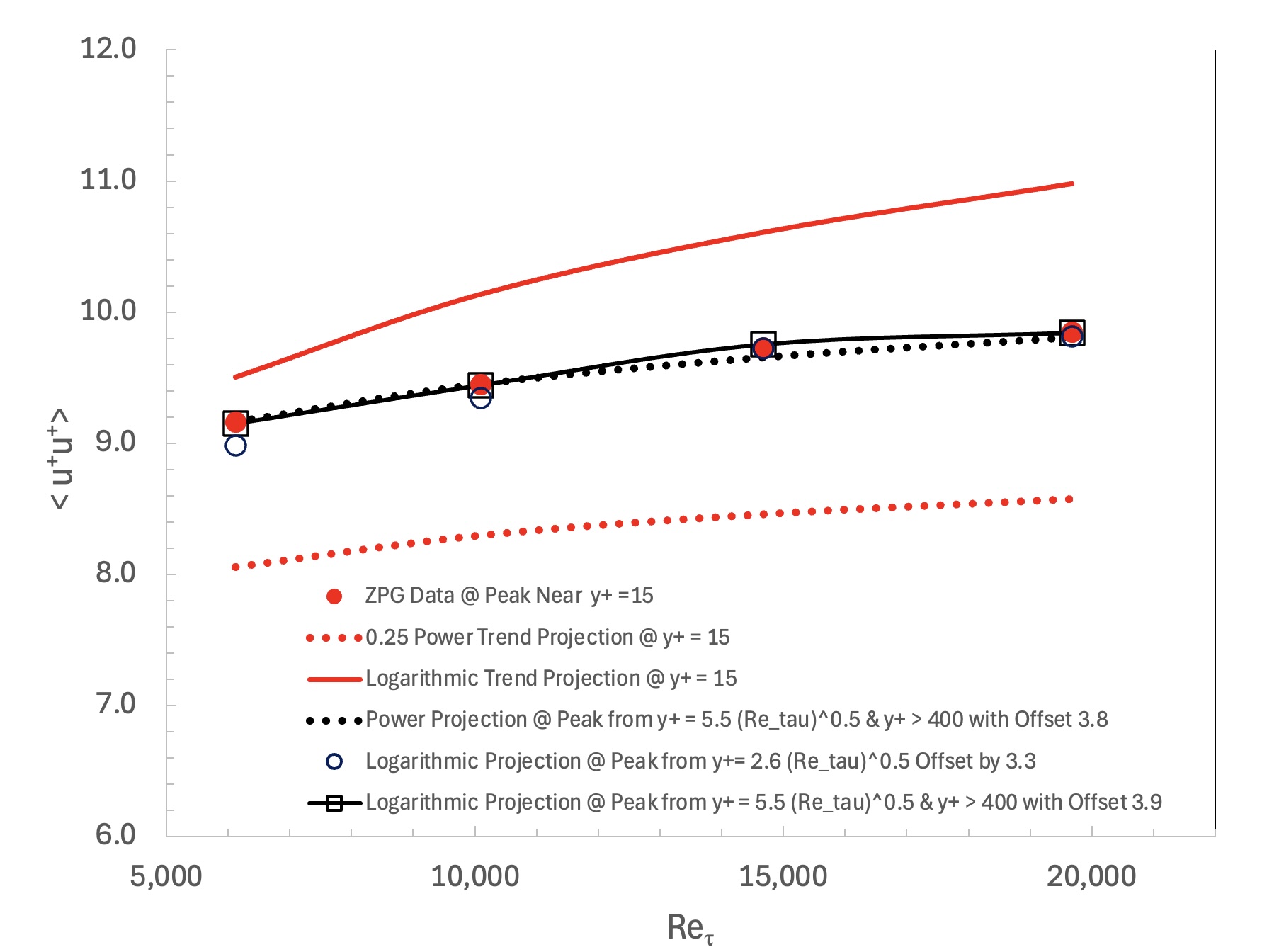}
    \caption{\label{fig:fig13} Values of streamwise normal stress at peak from ZPG data of \cite{sam18}, compared to various projections from logarithmic and power relations, including from fitting region at positions selected  using $430 < y^+ = 5.5 (Re_\tau)^{0.5} < 772$.}
     
\end{figure}

Careful examination of figures \ref{fig:fig1}, \ref{fig:fig2}, \ref{fig:fig3} and \ref{fig:fig5} suggests that the fitting region where the two models may be applicable, starts a distance from the wall $y^+_{in} \sim (Re_\tau)^{0.5}$;  $y^+_{in} \gtrapprox 400$ for the range of Reynolds numbers of the ZPG data. The outer limit of this fitting region scales with outer-scaled distance $Y$. From figure 6 of \cite{nag24}, and recalling that $y^+ = Y \cdot Re_\tau$, we tested several coefficients for the dependence of the lower limit of the fitting region on $(Re_\tau)^{0.5}$. 
First, we used the best fit of each model, with the data in the fitting region, to project down to $y^+ = 15$ to obtain the two red lines in figure \ref{fig:fig13}. It is clear that neither model directly represents the measured conditions around the inner peak values. However, when the projection is based on the normal stress value from a wall distance that varies with $(Re_\tau)^{0.5}$ and an offset is selected to match the highest $Re_\tau$ data point, and utilized for all Reynolds numbers, the respective projections for both models are the black open symbols and the dotted black line. For the logarithmic model, we initially tried a coefficient of $2.6$ for the $Re_\tau^{0.5}$ relation from the results of \cite{Klew09} based on channel DNS data of limited Reynolds numbers that provided a range $203 < y^+_{in} < 365$, which is not within the limits of the considered fitting region.  We found a value of $5.5$ with a larger offset to produce slightly better agreement with experimental data at higher Reynolds numbers in ZPG boundary layers based on  $y^+_{in} > 400$. In summary, Figure \ref{fig:fig13} demonstrates that the measured ZPG data of \cite{sam18} display a trend of the peak of streamwise normal stress with $Re_\tau$ comparable to that readily projected by the trend of the logarithmic relation and slightly faster than that of the $0.25$-power trend.

\begin{figure}
       \centering
      \includegraphics[width=0.95\textwidth]{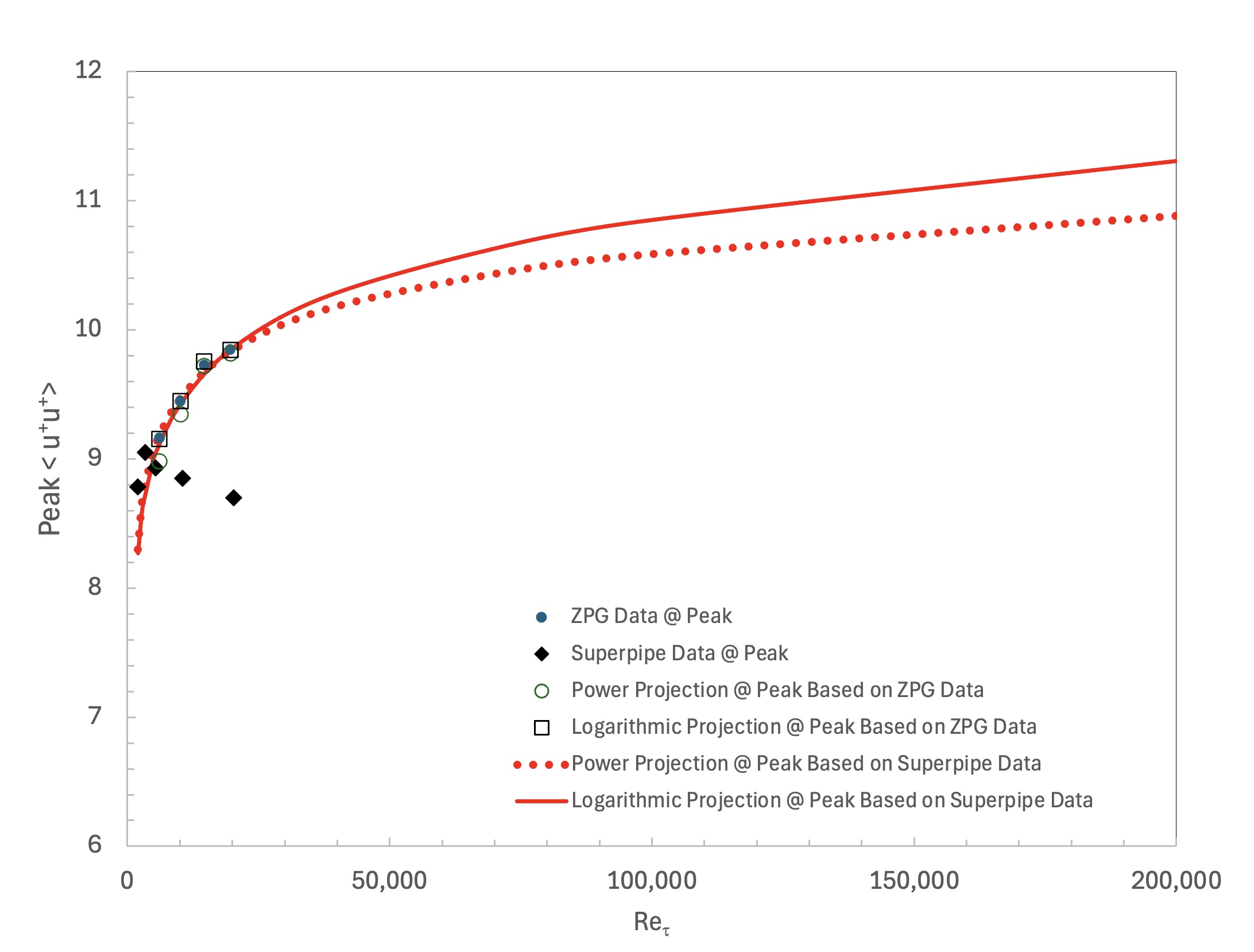}
    \caption{\label{fig:fig14}  Lower-$Re_\tau$ normal stress data measured in the Superpipe and used to project the trends by both models to higher $Re_\tau$ conditions achievable in facility using an offset of 4.2 for logarithmic trend and 3.8 for power trend. ZPG data from figure \ref{fig:fig13} are included to demonstrate near wall similarity.}
     
\end{figure}

The excellent agreement between the measured streamwise normal stress, around $y^+ \approx 15$ from ZPG data of \cite{sam18} with projections at the same peak position, based on data from locations at $y^+ = 5.5~(Re_\tau)^{0.5}$, which are in the range $430 < y^+ < 772$, supports the concept of influence by the outer flow on the inner peak growth with $Re_\tau$.  The approach was also used to demonstrate the potential of predicting the degree of that influence by extending the Superpipe peak measurements to higher $Re_\tau$ values. Again, selecting the locations in the data to project using the same approach with data at $y^+ = 5.5~(Re_\tau)^{0.5}$, which are in the range $400 < y^+ < 1,800$, but adjusting the offset based on the few more reliable data at lower $Re_\tau$ from the Superpipe, we obtain the red lines of figure \ref{fig:fig14}.  From a figure for the Superpipe data similar to figure \ref{fig:fig13}, we find that the average difference in the streamwise normal stress values at the inner peak around $y^+ = 15$ based on the two models is $30 \%$.  When each model trend is anchored by experimental data utilizing the approach used in figure \ref{fig:fig13}, a difference between the two relations at $Re_\tau$ of $10^6$ is estimated to be approximately $6.2 \%$, which is very challenging to accurately discriminate by measurements, especially at such high $Re_\tau$ values; see figure \ref{fig:fig14}, \cite{mar17} and \cite{nag23}.

Comparisons of experimental data for ZPG by \cite{sam18} with DNS data for channel flow by \cite{lee15}, coupled with projections for the peak normal stress values based on ``fitting region'' values are shown in figure \ref{fig:fig15}. The results reveal that high Reynolds number behavior is only achieved at around $Re_\tau$ of $5,000$. The results also indicate that at lower $Re_\tau$, the projected power trend at the peak is slightly more representative of the DNS data. For higher $Re_\tau$, predictions of the peak value of streamwise normal stress by both the logarithmic and power trends are equally representative.

\section{Conclusions}
To extend the work of \cite{nag24} to higher $Re_\tau$'s in wall-bounded flows, it was necessary to develop a new method to evaluate two models with proposed trends for streamwise normal stress using experimental results.  Two of the best available experimental data sets in ZPG boundary layers and pipe flows were identified, and provided a wide range of Reynolds numbers with $2,500 \lessapprox Re_\tau \lessapprox 20,000$ in ZPG boundary layers and $3,000 \lessapprox Re_\tau \lessapprox 100,000$ for fully-developed pipe flows. 

We find that to establish a significant region of validity by either relation, $Re_\tau \gtrapprox 10,000$. The two parameters for both models of equations \ref{eq:001} and \ref{eq:002} are found to be similar for the two flows, and using the same parameters for full range of $Re_\tau$ in each flow, results in a lower limit of wall distance independent of Reynolds number. For both flows and models, the lower limit in distance to the wall is $y^+_{in} \gtrapprox 400$, which corresponds to  $0.004 \lessapprox Y_{in} \lessapprox 0.04$ for the range of Reynolds numbers for both measurements in the ZPG boundary layer and the Superpipe examined here. However, the outer limit of validity for the power trend is nearly double that of the logarithmic trend, and is identified at $Y_{out} \approx 0.39$ versus $\approx0.22$ for ZPG, and $Y_{out} \approx 0.5$ versus $\approx0.27$ for Superpipe data. Both models are not valid outside of this fitting region identified between $y^+_{in}$ and $Y_{out}$. We also find that the standard deviation from the mean values of the two parameters for each of the proposed relations was quite small for all the data analyzed, and did not exceed $2.5\%$ of the mean values of the respective parameters; see captions of figures \ref{fig:fig1} through \ref{fig:fig10}.

\begin{figure}
      \centering
      \includegraphics[width=0.95\textwidth]{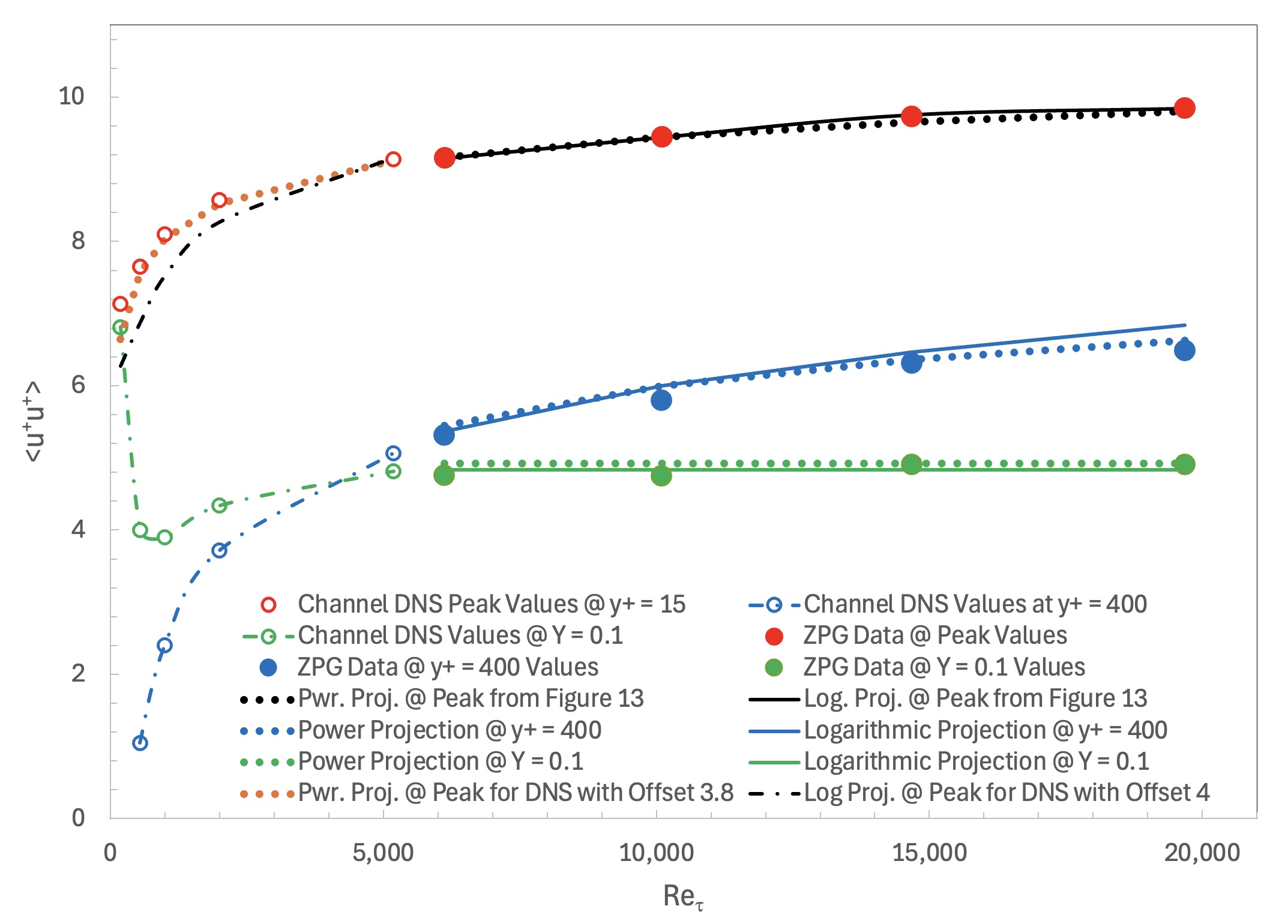}
    \caption{\label{fig:fig15} Comparison of projections by both models of peak streamwise normal stress values based on data from channel DNS of \cite{lee15} and higher $Re_\tau$ data from ZPG experiments by \cite{sam18}, as developed in figure \ref{fig:fig13}, including from outer part of DNS using $67<y^+=5.5(Re_\tau)^{0.5}<360$ and at $y^+=400$ and $Y=0.1$.}
     
\end{figure}

Interestingly, a somewhat larger exponent for the power law of streamwise normal stress equal to $0.28$, instead of $0.25$ obtained by the bounded dissipation assumption of \cite{che22} and \cite{che23}, is found to  extend its range of validity. Also, recognizing that the outer part of boundary layers is intermittent between laminar and turbulent conditions,  a correction to the normal stress is applied by dividing it by the intermittency factor as representative of the fraction of time the flow is turbulent.  The resulting validity of the the power trend is extended from $Y_{out} \approx 0.4$ to around half the boundary layer thickness. 

While it was not part of the initial objectives of the work, we attempted to project the expected normal stress at the inner peak around $y^+ \approx 15$ using normal stress values from within the fitting region to the right of the curved gray line of figures \ref{fig:fig1} and \ref{fig:fig2}. 
This approach can be beneficial as it leverages more easily and accurately measured normal stress values in wall-bounded flows. We relied on both models examined here, using the parameters established for each, and successfully tested the approach in the ZPG boundary layers. Applying this approach to the limited range of inner-peak measurements from the Superpipe, we project the expected peak normal stress up to $Re_\tau = 200,000$, covering nearly the full range of available conditions in the facility, as shown in figure \ref{fig:fig14}. The differences between the projections by the two models are small over the available range of data, as pointed out by \cite{nag23}.  When the trends of the models are anchored by experimental data, even at the higher $Re_\tau$ conditions typical of neutral atmospheric boundary layers around $Re_\tau = 10^6$, the two relations provide projected values with a relative difference approximately equal to $6.2\%$.  Therefore, to differentiate in favor of either would require measurement accuracy beyond our reach.

The excellent comparison of channel DNS data of \cite{lee15} up to $Re_\tau = 5,200$ and the higher Reynolds numbers data in ZPG boundary layers by \cite{sam18} for the peak streamwise normal stress, along with projections by both models, in figure \ref{fig:fig15} is encouraging. These results are supportive of the recommendations in \cite{hoy24b} and \cite{nag24}, that well resolved and converged DNS data in fully developed duct flows with adequate domain size up to around $Re_\tau = 5,000$, may be sufficient and more valuable to obtain accurate mean flow and turbulence statistics. DNS at higher Reynolds numbers may only be required for studies of structure and scales of turbulence, but they also need to be well resolved and converged and in long domains. The smooth continuation between the channel DNS data and the experimental ZPG boundary layer measurements for the peak of the normal stresses in figure \ref{fig:fig15} is further confirmation of the universality of the near wall region across different wall-bounded flows (pipes, channels, boundary layers, etc.).  The channel DNS data at $Y = 0.1$ for the lowest two $Re_\tau$ values reflect the very low Reynolds number conditions of $182$ and $543$ where an overlap is not sufficiently established as for $Re_\tau \gtrapprox 1,000$.

Finally, while both models may be used to represent the general trends of the streamwise normal-stress data, the power relation conforms to the experimental data more closely throughout its wider range of validity as demonstrated in figures \ref{fig:fig1} and \ref{fig:fig2}. The current results and those of \cite{mon23} and \cite{bax24} on the mean flow, revealing a linear term of the same order as the logarithmic term in the fitting region of the mean velocity profile, suggest that beyond the inner flow, $y^+ \gtrapprox 400$, consideration of nonlinear scale growth from the wall and accounting for some viscous effects are important for modeling overlap/inertial region of wall-bounded flows.  Such effects also apply to potential refinements of the logarithmic trend along ideas advanced by \cite{Desh21}.

\section*{Acknowledgement}

HN acknowledges the support of the Rettaliata Chair Professorship at ILLINOIS TECH and accommodations provided by University of Melbourne during a short sabbatical period when the authors worked to complete this work. IM received financial support from the Office of Naval Research (ONR) through ONR global grant: N62909-23-1-2068. The support of the King Abdullah University of Science and Technology, Saudi Arabia, is also acknowledged for organizing a three-day workshop at KAUST on ``Outstanding challenges in wall turbulence and lessons to be learned from pipes", from Monday 26 to Wednesday 28 February 2024, where the interaction between the authors on this topic was energized and led to the work included here. 

\bibliographystyle{jfm_old}
\bibliography{Normal_stress_trends}

\end{document}